\documentclass[english,aps,preprint]{revtex4}
\usepackage[T1]{fontenc}
\usepackage[latin1]{inputenc}
\usepackage{graphicx}
\makeatletter

\usepackage{babel}
\makeatother
\begin{document}
\title{Statistics of Semiflexible Polymer Chains and the Generalized Borel Transform}

\author{Marcelo Marucho }

\email{marucho@polymer.uakron.edu}

\author{Gustavo A. Carri%
\footnote{To whom any correspondence should be addressed%
}}

\email{carri@polymer.uakron.edu}

\affiliation{The Maurice Morton Institute of Polymer Science, The University of
Akron, Akron, OH 44325-3909, USA.}

\begin{abstract}
In this paper, we present a new approach to the discrete version of the Wormlike Chain Model (WCM) of semiflexible polymers. Our solution to the model is based on a new computational technique called the Generalized
Borel Transform (GBT) which we use to study the statistical mechanics of semiflexible polymer chains.
Specifically, we evaluate the characteristic function of the model approximately. Afterward, we compute the polymer
propagator of the model using the GBT and find an expression valid
for polymers with any number of segments and values of the semiflexibility
parameter. This expression captures the limits of flexible and infinitely
stiff polymers exactly. In between, a smooth and approximate crossover
behavior is predicted. Another property of our propagator is that it
fulfills the condition of finite extensibility of the polymer chain.
We have also calculated the single chain structure factor. This property
is a decreasing function of the wave vector, $k,$ until a plateau is
reached. Our computations clearly show that the structure factor decreases
faster with increasing wave vector when the semiflexibility parameter
is increased. Furthermore, when the wave vector is large enough, there
is a regime where the structure factor follows an approximate power law
of the form $k^{-\theta}$ even for short polymer chains. $\theta$
is equal to two for flexible polymers and to one for rigid chains. We also compare our results to the predictions of other models.
\end{abstract}
\maketitle

\pagebreak

\section{Introduction}

In recent years, experimental studies of biological macromolecules
have motivated intense research in the field of statistical mechanics
of single semiflexible polymer chains. Indeed, studies like force-elongation
measurements of different biological (DNA, titin, tenascin) and synthetic
(polyethylene glycol, polyvinyl alcohol) polymers using Atomic Force
Microscopes, Optical Tweezers and other recently developed tools are
abundant\cite{force}. The measured force-elongation curve is generally
fitted to the prediction of the Wormlike Chain Model (WCM) of semiflexible
chains\cite{Marko}, originally proposed by Kratky and Porod\cite{Porod}.
From this fit, parameters like the persistence length of the biopolymer
are extracted. Another kind of experiments has targeted the mechanical
properties of eukaryotic cells\cite{expe}. These properties are determined
by an assembly of protein fibers called the cytoskeleton. This three
dimensional assembly is made of the cytoskeletal polymers (microtubules,
actin filaments, etc.). All these polymers are semiflexible polymers
at the relevant length scales (a few microns at most). Thus, the predictions
of the WCM are very relevant for the understanding of the physical
behavior of cytoskeletal polymers. 

The WCM was originally proposed by Kratky and Porod in 1949\cite{Porod}
and reformulated using field theoretic methods by Saito \textit{et
al}.\cite{saito} in 1967. In this model, the polymer chain displays
resistance to bending deformations. This resistance is modeled using
a free energy that penalizes bending the polymer backbone. The free
energy depends on parameters (elastic constants) that are a consequence
of many short-range monomer-monomer interactions. Explicitly, the
free energy is

\begin{equation}
\frac{\kappa}{2}\int_{0}^{L}\mathrm{ds}\left(\frac{\mathrm{d}^{2}\mathbf{R}\left(\mathrm{s}\right)}{\mathrm{ds}^{2}}\right)^{2},\label{eq:fffdd}\end{equation}
where $\mathbf{R}\left(\mathrm{s}\right)$ is the vectorial field
that represents the polymer chain, $s$ is the arc of length parameter,
$L$ is the contour length of the polymer and $\kappa$ is the bending
modulus. In addition, the local inextensibility constraint $\left|\mathrm{d}\mathbf{R}\left(\mathrm{s}\right)/\mathrm{ds}\right|=1$
must be satisfied. As a consequence of the bending rigidity , a wormlike
chain is characterized by a persistence length (proportional to the
bending modulus) such that, if the length scale is shorter than the
persistence length, then the chain behaves like a rod while, if the
length scale is larger than the persistence length, then the chain
is governed by the configurational entropy that favors the random-walk
conformations. 

The local inextensibility constraint has not allowed researchers to
find an exact solution to the WCM. Indeed, the constraint $\left|\mathrm{d}\mathbf{R}\left(\mathrm{s}\right)/\mathrm{ds}\right|=1$
is written using a Dirac delta distribution in infinite dimensions.
Depending on how the constraint is written, $\left|\mathrm{d}\mathbf{R}\left(\mathrm{s}\right)/\mathrm{ds}\right|=1$
or $\left(\mathrm{d}\mathbf{R}\left(\mathrm{s}\right)/\mathrm{ds}\right)^{2}=1$,
we get an Edwards Hamiltonian that is non-analytic or non-linear,
respectively. Consequently, there is no exact solution of this model
at present. However, a few properties like the first few moments of
the distribution of the end-to-end distance\cite{saito,Yamakawa_2}
are known exactly.

The aforementioned complexity of the WCM has motivated many approximate
treatments of semiflexible polymers. For example, Fixman and Kovac\cite{Fixman}
developed a modified Gaussian model for stiff polymer chains under
an external field (external force). In this approach, they computed
an approximate distribution for the bond vectors from which they were
able to compute the partition function and average end-to-end vector.
An alternative approach was proposed by Harris and Hearst\cite{Harris}
who developed a distribution for the continuous model from which they
were able to compute the two-point correlation function and, consequently,
the mean-square end-to-end distance and radius of gyration. Another
statistical property of interest has been the distribution function
of the end-to-end distance or its Fourier transform. Many approximations
for this function have been proposed. For example, different expansions
of the distribution in inverse powers of the number of segments have
been developed\cite{Daniels,Yama}. Similarly, perturbations with
respect to the rodlike limit have been derived\cite{Frey}. Other
approaches to the distribution function of the end-to-end distance
of semiflexible polymers have led to modified Gaussian functions\cite{Baw,Harnau}.
Finally, many approximations have been proposed for the structure
factor\cite{Marques,struc}. 

As stated in the previous paragraph, most of the approximate treatments
of the distribution function of the end-to-end distance have been
perturbative in nature. Indeed, these approximations were perturbation
expansions with respect to the flexible or rigid chain limits. A different approach to semiflexible polymers was taken by Kholodenko\cite{KK,Kholodenko,More-Kholodenko}. In this model, the Euclidean version of the Dirac propagator is used to predict the conformational properties of semiflexible polymers. In particular, the single chain structure factor has been used to describe experimental data quantitatively\cite{Ballauf}. Recently, Winkler has proposed another treatment of semiflexible polymers\cite{Winkler}. In this work, an approximate expression for the distribution function valid for
any value of the stiffness of the polymer backbone was developed
using the Maximum Entropy Principle.

Motivated by the experimental studies done on semiflexible polymers
and our incomplete understanding of the properties of the WCM, we
have developed a new approach that captures many physical properties
of the model, like the limits of flexible and rigid polymers, exactly
and provides approximate crossover behaviors for all the distribution
functions. To accomplish this goal, we have employed a computational
technique called the Generalized Borel Transform (GBT) which was taken
from Quantum Mechanics and Quantum Field Theories\cite{tecnica1,tecnica2,tecnica3}.
This method computes Mellin/Laplace transforms exactly. We provide
a brief summary of this technique in Appendix A. 

This paper is organized as follows. In section II, we evaluate the
characteristic function of the WCM approximately such that some physical
constraints are satisfied exactly. Furthermore, we evaluate the distribution
function (polymer propagator) using the GBT and compute the single
chain structure factor. In Section III we discuss the results of our
calculations which are valid for any value of the semiflexibility
of the polymer. Section IV contains the conclusions of our work and
some speculations about applications to polymer physics of possible
extensions of the GBT. The details of the mathematical calculations
are presented in the appendices.

\section{theory}

\subsection{The Model and Evaluation of the Characteristic Function}

Consider a polymer chain modeled as a sequence of $n$ bond vectors
$\left(\mathbf{R}_{1},\mathbf{R}_{2},...,\mathbf{R}_{n}\right)$ connected
in a sequential manner. In addition, let us assume that the length
of each bond vector is $l$ (=Kuhn length) and that pairs of consecutive
bond vectors try to be parallel to each other. This preferential orientation
is modeled with a Boltzmann weight given by the following expression
\cite{Fixman,batha,Harnau,Marques}

\begin{equation}
{\displaystyle \exp\left(\alpha\sum_{i=1}^{n-1}\mathbf{R}_{i+1}\cdot\mathbf{R}_{i}\right)},\label{eq:potential_energy}\end{equation}
where $\alpha$ is the strength of the interaction in units of thermal
energy $\left(=k_{B}\, T\right)$. Inserting Eq. (\ref{eq:potential_energy})
into the expression for the propagator of the Random Flight Model\cite{Karl},
we obtain the following expression for the polymer propagator of semiflexible
chains 

\begin{equation}
{\displaystyle P\left(\mathbf{R},n,\alpha\right)=\int d\left\{ \mathbf{R}_{k}\right\} \prod_{j=1}^{n}\tau\left(\mathbf{R}_{j}\right)\delta\left(\sum_{j=1}^{n}\mathbf{R}_{j}-\mathbf{R}\right)\exp\left(\alpha\sum_{i=1}^{n-1}\mathbf{R}_{i+1}\cdot\mathbf{R}_{i}\right)},\label{eq:definition}\end{equation}
where $\mathbf{R}$ is the end-to-end vector and $\tau\left(\mathbf{R}_{j}\right)$
is given by the formula 

\begin{equation}
{\displaystyle \tau\left(\mathbf{R}\right)=\frac{\delta\left(\left|\mathbf{R}\right|-l\right)}{4\pi l^{2}}}.\label{eq:constraint}\end{equation}
The propagator, Eq. (\ref{eq:definition}), is not normalized. 

We proceed to express the delta function using its Fourier representation\cite{Zinn-Justin}
then, Eq. (\ref{eq:definition}) becomes

\begin{equation}
\begin{array}{c}
\begin{array}{c}
P\left(\mathbf{R},n,\alpha\right)={\displaystyle \int{\displaystyle \frac{d^{3}k\exp\left(-i\mathbf{R}\cdot\mathbf{k}\right)}{\left(2\pi\right)^{3}\left(4\pi l^{2}\right)^{n}}}\left[\int d\left\{ \mathbf{R}_{k}\right\} \prod_{j=1}^{n}\delta\left(\left|\mathbf{R}_{j}\right|-l\right)\right.}\\
\;\end{array}\\
{\displaystyle \times\left.\exp\left({\displaystyle \mathrm{i}\sum_{j=1}^{n}\mathbf{R}_{j}\cdot\mathbf{k}\mathrm{+}\alpha\sum_{i=1}^{n-1}\mathbf{R}_{i+1}\cdot\mathbf{R}_{i}}\right)\right]},\end{array}\label{eq:propagator_Fourier}\end{equation}
which can be used to define the characteristic function, $K\left(\mathbf{k},\alpha,n,l\right)$,
as follows

\begin{equation}
{\displaystyle P\left(\mathbf{R},n,\alpha\right)\equiv{\displaystyle \int\frac{d^{3}k}{\left(2\pi\right)^{3}}\exp\left(-i\mathbf{R}\cdot\mathbf{k}\right)K\left(\mathbf{k},\alpha,n,l\right)}}.\label{eq:distri}\end{equation}
The mathematical expression of the characteristic function is

\begin{equation}
{\displaystyle K\left(\mathbf{k},\alpha,n,l\right)\equiv\frac{1}{\left(4\pi l^{2}\right)^{n}}\int d\left\{ \mathbf{R}_{k}\right\} \prod_{j=1}^{n}\delta\left(\left|\mathbf{R}_{j}\right|-l\right)\exp\left(i\sum_{j=1}^{n}\mathbf{R}_{j}\cdot\mathbf{k}+\alpha\sum_{i=1}^{n-1}\mathbf{R}_{i+1}\cdot\mathbf{R}_{i}\right)}.\label{eq:caracprim}\end{equation}
Note that since the polymer propagator, Eq. (\ref{eq:definition}),
is not normalized then the characteristic function, Eq. (\ref{eq:caracprim}),
does not approach one when the wave vector goes to zero. Instead,
it approaches the canonical partition function of the model. Furthermore,
note that the characteristic function is a Fourier Transform in a
$3n$-dimensional space.

As stated before by Yamakawa\cite{Yamakawa_2}, the exact evaluation
of the characteristic function (or the polymer propagator) for semiflexible
chains is not possible at present. Therefore, we have developed a
new approximation to evaluate this function. This new mathematical
approach was developed in such a way that the most relevant physics
of the problem is not altered by the approximation. Specifically,
the proposed approach keeps the thermodynamics (partition function)
of the model exact. Moreover, all the properties of fully flexible
chains $\left(\alpha\rightarrow0\right)$ and infinitely stiff chains
$\left(\alpha\rightarrow\infty\right)$ are preserved exactly. Consequently,
this approach captures both asymptotic limits exactly and provides
an approximate description of the crossover behavior. In addition,
our treatment of the problem uses the exact expression of the mean-square
end-to-end distance. Consequently, this quantity and the mean-square
radius of gyration are exact. Another important property of our approach
is that it keeps the local inextensibility constraint intact. Therefore,
our chains have finite extensibility and this model can be used to
compute the force-elongation relationship of semiflexible polymers.
We describe our approximation hereafter.

Let us start by computing the following class of integrals 

\begin{equation}
{\displaystyle G_{j}=\int d\mathbf{R}_{j}\delta\left(\left|\mathbf{R}_{j}\right|-l\right)\exp\left(i\mathbf{R}_{j}\cdot\mathbf{k}+\alpha\mathbf{R}_{j+1}\cdot\mathbf{R}_{j}\right)},\label{eq:Gj}\end{equation}
which are present in the characteristic function. The wave vector
$\mathbf{k}$ is constant and can be chosen in the direction of the
versor $\hat{\mathbf{z}}$. Writing all the vectors in spherical coordinates
we can express $G_{j}$ as follows

\begin{equation}
\begin{array}{c}
{\displaystyle G_{j}=\int d\left(\left|\mathbf{R}_{j}\right|\right)\left|\mathbf{R}_{j}\right|^{2}\delta\left(\left|\mathbf{R}_{j}\right|-l\right)\int_{0}^{\pi}d\theta_{j}\sin\left(\theta_{j}\right)\exp\left\{ i\left|\mathbf{R}_{j}\right|\cdot\left|\mathbf{k}\right|\cos\left(\theta_{j}\right)\right\} }\\
{\displaystyle \times\exp\left\{ \alpha\left|\mathbf{R}_{j}\right|\cdot\left|\mathbf{R}_{j+1}\right|\cos\left(\theta_{j+1}\right)\cos\left(\theta_{j}\right)\right\} \int_{0}^{2\pi}d\varphi_{j}\exp\left(\gamma\cos\left(\varphi_{j}\right)+\beta\sin\left(\varphi_{j}\right)\right),}\end{array}\label{eq:Gjs}\end{equation}
where $\gamma$ and $\beta$ are defined as follows

\begin{equation}
\begin{array}{c}
{\displaystyle \gamma\equiv\alpha\left|\mathbf{R}_{j}\right|\left|\mathbf{R}_{j+1}\right|\sin\left(\theta_{j}\right)\cos\left(\varphi_{j+1}\right)\sin\left(\theta_{j+1}\right),}\\
{\displaystyle \beta\equiv\alpha\left|\mathbf{R}_{j}\right|\left|\mathbf{R}_{j+1}\right|\sin\left(\theta_{j}\right)\sin\left(\varphi_{j+1}\right)\sin\left(\theta_{j+1}\right).}\end{array}\label{eq:iuty}\end{equation}

The $\varphi_{j}$-integrals can be done exactly. The result is 

\begin{equation}
{\displaystyle F_{j}\left(\gamma,\beta\right)=\int_{0}^{2\pi}d\varphi_{j}\exp\left(\gamma\cos\left(\varphi_{j}\right)+\beta\sin\left(\varphi_{j}\right)\right)=2\pi I_{0}\left(\alpha\left|\mathbf{R}_{j}\right|\left|\mathbf{R}_{j+1}\right|\left|\sin\left(\theta_{j}\right)\sin\left(\theta_{j+1}\right)\right|\right)},\label{eq:Fj}\end{equation}
where $I_{0}\left(x\right)$ is the Bessel function of second class\cite{abramovich}.
After we integrate the delta function, the function $G_{j}$ becomes

\begin{equation}
{\displaystyle G_{j}=2\pi l^{2}\int_{0}^{\pi}d\theta_{j}\sin\left(\theta_{j}\right)\exp\left\{ \left[il\cdot\left|\mathbf{k}\right|+\alpha l^{2}\cos\left(\theta_{j+1}\right)\right]\cos\left(\theta_{j}\right)\right\} I_{0}\left(\alpha l^{2}\left|\sin\left(\theta_{j}\right)\sin\left(\theta_{j+1}\right)\right|\right).}\label{eq:Gj3}\end{equation}

We now replace this expression into Eq. (\ref{eq:caracprim}) and
obtain the characteristic function 

\begin{equation}
\begin{array}{c}
{\displaystyle K\left(\mathbf{k},\alpha,n,l\right)=2^{-n}\int_{0}^{\pi}d\theta_{n}{\displaystyle \sin\left(\theta_{n}\right)\exp\left\{ il\cdot\left|\mathbf{k}\right|\cos\left(\theta_{n}\right)\right\} \int_{0}^{\pi}\prod_{j=1}^{n-1}d\theta_{j}\sin\left(\theta_{j}\right)}}\\
\times{\displaystyle \exp\left\{ \left[il\cdot\left|\mathbf{k}\right|+\alpha l^{2}\cos\left(\theta_{j+1}\right)\right]\cos\left(\theta_{j}\right)\right\} I_{0}\left(\alpha l^{2}\left|\sin\left(\theta_{j}\right)\sin\left(\theta_{j+1}\right)\right|\right)}.\end{array}\label{eq:charactI}\end{equation}

The evaluation of $K\left(\mathbf{k},\alpha,n,l\right)$ is done by
iterations. First, we take the term $j=1$, redefine $\left|\mathbf{k}\right|\rightarrow k/l$
and $\alpha\rightarrow\alpha/l^{2}$, and remove the factor $2 \pi l^2$ from the definition of $G_j \left( \theta \right)$ in Eq. (\ref{eq:Gj3}) ; then, we can write

\begin{equation}
{\displaystyle G_{1}\left(\theta_{2}\right)=\int_{0}^{\pi}d\theta_{1}\sin\left(\theta_{1}\right)\exp\left\{ \left[ik+\alpha\cos\left(\theta_{2}\right)\right]\cos\left(\theta_{1}\right)\right\} I_{0}\left(\alpha\left|\sin\left(\theta_{1}\right)\sin\left(\theta_{2}\right)\right|\right)},\label{eq:geuno}\end{equation}
which is exactly doable\cite{tablarusa}. The result is

\begin{equation}
{\displaystyle G_{1}\left(\theta_{2}\right)=2\frac{\sinh\sqrt{\alpha^{2}-k^{2}+2ik\alpha\cos\left(\theta_{2}\right)}}{\sqrt{\alpha^{2}-k^{2}+2ik\alpha\cos\left(\theta_{2}\right)}}}.\label{eq:G1final}\end{equation}

The next step in the iterative process is the evaluation of $G_{2}\left(\theta_{3}\right)$
given by 

\begin{equation}
{\displaystyle G_{2}\left(\theta_{3}\right)=\int_{0}^{\pi}d\theta_{2}\sin\left(\theta_{2}\right)\exp\left\{ \left[ik+\alpha\cos\left(\theta_{3}\right)\right]\cos\left(\theta_{2}\right)\right\} I_{0}\left(\alpha\sin\left(\theta_{2}\right)\sin\left(\theta_{3}\right)\right)G_{1}\left(\theta_{2}\right)}.\label{eq:G2}\end{equation}
This integral is not exactly doable. Consequently, we proceed to approximate
it such that the asymptotic limits of flexible and stiff polymers
are captured exactly. Thus, the expression that we obtain will give
an approximate crossover behavior between the aforementioned limiting
regimes. Note that in the limit of very stiff chains, $\alpha\rightarrow\infty$,
all the segments will be parallel to each other. In other words, when
$\alpha\rightarrow\infty$, $\theta_{2}\simeq\theta_{3}$. Then, in
this limit we can say that 

\begin{equation}
{\displaystyle G_{2}\left(\theta_{3}\right)\simeq\left[G_{1}\left(\theta_{3}\right)\right]^{2}}.\label{eq:G2aprox}\end{equation}
In the other limit, $\alpha\rightarrow0$, $G_{1}\left(\theta_{2}\right)$
is independent of $\theta_{2}$. Therefore, Eq. (\ref{eq:G2aprox})
is also valid in the limit of flexible chains. Thus, we conclude that
Eq. (\ref{eq:G2aprox}) is a good approximation for $G_{2}\left(\theta_{3}\right)$
since it captures the asymptotic limits exactly and provides an approximate
crossover behavior for $G_{2}\left(\theta_{3}\right)$.

The iteration of the aforementioned approximation $n-1$ times leads
to the following expression for the characteristic function 

\begin{equation}
{\displaystyle K\left(k,\alpha,n\right)\simeq\frac{1}{2^n}\int_{0}^{\pi}d\theta_{n}\sin\left(\theta_{n}\right)\exp\left\{ ik\cos\left(\theta_{n}\right)\right\} \left[G_{1}\left(\theta_{n}\right)\right]^{n-1}}.\label{eq:chac}\end{equation}
Note that this expression gives the exact canonical partition function
of the model

\begin{equation}
{\displaystyle Z_{n}\left(\alpha\right)=K\left(0,\alpha,n\right)=\left(\frac{\sinh\alpha}{\alpha}\right)^{n-1}}.\label{eq:cano}\end{equation}
Thus, this first part of the approximation preserves both asymptotic
behaviors and the thermodynamics of the problem intact. 

Let us now proceed to evaluate the approximate expression of the characteristic
function, $K\left(k,\alpha,n\right)$. The integral in Eq. (\ref{eq:chac})
is not exactly doable thus, we evaluate it using a variational procedure.
Let us introduce the following anzats 

\begin{equation}
{\displaystyle \sqrt{\alpha^{2}-k^{2}+2ik\alpha\cos\left(\theta_{n}\right)}=\sqrt{\alpha^{2}-k^{2}\nu_{\alpha,n}^{2}}+ikg_{\alpha,n}\cos\left(\theta_{n}\right)},\label{eq:anzats}\end{equation}
where the parameters $g_{\alpha,n}$ and $\nu_{\alpha,n}$ are determined
from the constraints imposed by the physics of the problem as described
below. One of the requirements is that the flexible and rigid limits
are captured exactly by the model. This requires that the parameters
must behave in the following way

\begin{equation}
{\displaystyle \begin{array}{c}
g_{\alpha,n}\rightarrow0\\
\nu_{\alpha,n}^{2}\rightarrow1\end{array}\quad\alpha\rightarrow0,\qquad\begin{array}{c}
g_{\alpha,n}\rightarrow1\\
\nu_{\alpha,n}^{2}\rightarrow0\end{array}\quad\alpha\rightarrow\infty}.\label{eq:conditionasympt}\end{equation}

Using Eq. (\ref{eq:anzats}), we can approximate the characteristic
function as follows

\begin{equation}
\begin{array}{c}
{\displaystyle {\displaystyle K\left(k,\alpha,n\right)\simeq\int_{0}^{\pi}d\theta_{n}\sin\left(\theta_{n}\right)\exp\left\{ ik\cos\left(\theta_{n}\right)\left[1+\left(n-1\right)g_{\alpha,n}\right]\right\} }}\\
{\displaystyle \times\frac{\left[\exp\left\{ \sqrt{\alpha^{2}-k^{2}\nu_{\alpha,n}^{2}}\right\} -\exp\left\{ -\sqrt{\alpha^{2}-k^{2}\nu_{\alpha,n}^{2}}\left[1-\frac{{\displaystyle 2ikg_{\alpha,n}\cos\left(\theta_{n}\right)}}{{\displaystyle \sqrt{\alpha^{2}-k^{2}\nu_{\alpha,n}^{2}}}}\right]\right\} \right]^{n-1}}{2^{n}\left[\sqrt{\alpha^{2}-k^{2}\nu_{\alpha,n}^{2}}\right]^{n-1}\left[1+\frac{{\displaystyle ikg_{\alpha,n}\cos\left(\theta_{n}\right)}}{{\displaystyle \sqrt{\alpha^{2}-k^{2}\nu_{\alpha,n}^{2}}}}\right]^{n-1}}}\end{array}\label{eq:characapprox}\end{equation}
Note that the term

\begin{equation}
{\displaystyle ikg_{\alpha,n}\cos\left(\theta_{n}\right)/\sqrt{\alpha^{2}-k^{2}\nu_{\alpha,n}^{2}}},\label{eq:tira}\end{equation}
goes to zero as $\alpha^{-1}$ in the limit of $\alpha\rightarrow\infty$
and, when $\alpha\rightarrow0$, it also approaches zero because $g_{\alpha,n}\rightarrow0$.
Thus, neglecting this term does not alter the predictions of the model
for the flexible and stiff limits. Consequently, we approximate Eq.
(\ref{eq:characapprox}) as follows 

\begin{equation}
\begin{array}{c}
{\displaystyle K\left(k,\alpha,n\right)\simeq\frac{\left[\sinh\left\{ \sqrt{\alpha^{2}-k^{2}\nu_{\alpha,n}^{2}}\right\} \right]^{n-1}}{2\left[\sqrt{\alpha^{2}-k^{2}\nu_{\alpha,n}^{2}}\right]^{n-1}}}\\
{\displaystyle \times\int_{0}^{\pi}d\theta_{n}\sin\left(\theta_{n}\right)\exp\left\{ ik\cos\left(\theta_{n}\right)\left[1+\left(n-1\right)g_{\alpha,n}\right]\right\} },\end{array}\label{eq:characapprox1}\end{equation}
which is exactly doable. The final expression for the characteristic
function is 

\begin{equation}
{\displaystyle K\left(k,\alpha,n\right)\simeq\frac{\left[\sinh\left\{ \sqrt{\alpha^{2}-k^{2}\nu_{\alpha,n}^{2}}\right\} \right]^{n-1}}{\left[\sqrt{\alpha^{2}-k^{2}\nu_{\alpha,n}^{2}}\right]^{n-1}k\left[1+\left(n-1\right)g_{\alpha,n}\right]}\sin\left\{ k\left[1+\left(n-1\right)g_{\alpha,n}\right]\right\} .}\label{eq:chacfinal}\end{equation}

We note that the expression of the characteristic function given
by Eq. (\ref{eq:chacfinal}) recovers the exact expression of the
canonical partition function of the model, Eq. (\ref{eq:cano}), in
the limit of $k\rightarrow0$. 

Let us now proceed to determine the values of the parameters $\nu_{\alpha,n}$
and $g_{\alpha,n}$ from the physics of the problem. We first look
at the force-elongation behavior predicted by this model. This curve
is given by the following mathematical expression

\begin{equation}
{\displaystyle L=\frac{\partial\left\{ \textrm{ln}\left[K\left(iF,\alpha,n\right)\right]\right\} }{\partial F}=\frac{1}{K\left(iF,\alpha,n\right)}\frac{\partial\left\{ K\left(iF,\alpha,n\right)\right\} }{\partial F},}\label{eq:elongacionsemiflex}\end{equation}
where $F$ is the applied force and $L$ is the average end-to-end
distance of the polymer chain in the direction of the force. The physics
of the problem imposes the following constraint 

\begin{equation}
{\displaystyle \begin{array}{c}
\lim\\
F\rightarrow\infty\end{array}\frac{1}{K\left(iF,\alpha,n\right)}\frac{\partial\left\{ K\left(iF,\alpha,n\right)\right\} }{\partial F}=n,}\label{eq:constrain}\end{equation}
which represents the finite extensibility of the polymer chain. In
other words, the polymer chain cannot be stretched more than its total
contour length. This constraint, as expressed by Eq. (\ref{eq:constrain}),
results in the following relationship between the parameters $\nu_{\alpha,n}$
and $g_{\alpha,n}$

\begin{equation}
{\displaystyle g_{\alpha,n}=1-\nu_{\alpha,n},}\label{eq:secondfit}\end{equation}
which is in perfect agreement with the required asymptotic behaviors
given by Eq. (\ref{eq:conditionasympt}). 

Equation (\ref{eq:secondfit}) gives one of the two equations required
to determine the parameters $\nu_{\alpha,n}$ and $g_{\alpha,n}$
completely. The second equation is obtained from the mean-square end-to-end
distance, $\left\langle R^{2}\right\rangle _{\alpha,n}$. We require
that our approximation reproduce this statistical quantity exactly.
The exact mathematical expression of this average is \cite{Harnau,Benoit},

\begin{equation}
{\displaystyle \left\langle R^{2}\right\rangle _{\alpha,n}=\left[n\frac{1+\mathcal{L}\left(\alpha\right)}{1-\mathcal{L}\left(\alpha\right)}-2\mathcal{L}\left(\alpha\right)\frac{1-\mathcal{L}\left(\alpha\right)^{n}}{\left(1-\mathcal{L}\left(\alpha\right)\right)^{2}}\right]},\label{eq:rcuadrado}\end{equation}
where $\mathcal{L}\left(\alpha\right)$ is the Langevin function \cite{tablarusa}. 

In order to derive the second relationship between $\nu_{\alpha,n}$
and $g_{\alpha,n}$, we divide the characteristic function, Eq. (\ref{eq:chacfinal}),
by the canonical partition function and expand this ratio in powers
of the wave vector $k$ to second order. The result is the following

\begin{equation}
{\displaystyle \left\langle R^{2}\right\rangle _{\alpha,n}=-\frac{1}{Z_{n}\left(\alpha\right)}\bigtriangledown_{\mathbf{k}}^{2}\left.K\left(k,\alpha,n\right)\right|_{\mathbf{k}=0}=\left[1+\left(n-1\right)g_{\alpha,n}\right]^{2}+\frac{3\left(n-1\right)\nu_{\alpha,n}^{2}}{\alpha}\mathcal{L}\left(\alpha\right),}\label{eq:firstfit}\end{equation}
which completes our approximation. Equations (\ref{eq:secondfit}),
(\ref{eq:rcuadrado}) and (\ref{eq:firstfit}) determine $\nu_{\alpha,n}$
and $g_{\alpha,n}$ completely. Furthermore, the use of the exact
expression for $\left\langle R^{2}\right\rangle _{\alpha,n}$ assures
that our approximation predicts not only $\left\langle R^{2}\right\rangle _{\alpha,n}$
exactly, but also $\left\langle R_{g}^{2}\right\rangle _{\alpha,n}$
since they are related by the equation\cite{Benoit}

\begin{equation}
{\displaystyle R_{g}^{2}=\frac{1}{\left(n+1\right)^{2}}\sum_{ni=1}^{n}\left(n-ni+1\right)\left\langle R^{2}\right\rangle _{\alpha,ni}.}\label{eq:relargr}\end{equation}
The final expression for $\nu_{\alpha,n}$ is

\begin{equation}
{\displaystyle \nu_{\alpha,n}=\frac{2n\left(n-1\right)-\sqrt{4n^{2}\left(n-1\right)^{2}-4\left[\left(n-1\right)^{2}+\frac{3\left(n-1\right)}{\alpha}\mathcal{L}\left(\alpha\right)\right]\left(n^{2}-\left\langle R^{2}\right\rangle _{\alpha,n}\right)}}{2\left[\left(n-1\right)^{2}+\frac{3\left(n-1\right)}{\alpha}\mathcal{L}\left(\alpha\right)\right]},}\label{eq:nu}\end{equation}
and our approximation is complete. 

\subsection{Evaluation of the Polymer Propagator using the GBT}

Replacing the expression given by Eq. (\ref{eq:chacfinal}) into Eq.
(\ref{eq:distri}) we obtain the following approximate expression
for the polymer propagator

\begin{equation}
{\displaystyle P\left(R,n,\alpha\right)=\int_{0}^{\infty}dk\frac{\sin\left(kR\right)\sin\left\{ k\left[1+\left(n-1\right)g_{\alpha,n}\right]\right\} \left[\sinh\left\{ \sqrt{\alpha^{2}-k^{2}\nu_{\alpha,n}^{2}}\right\} \right]^{n-1}}{2\pi^{2}R\left[1+\left(n-1\right)g_{\alpha,n}\right]\left[\sqrt{\alpha^{2}-k^{2}\nu_{\alpha,n}^{2}}\right]^{n-1}}}.\label{eq:dis}\end{equation}

In the limit of $\alpha\rightarrow0$, Eq. (\ref{eq:dis}) becomes 

\begin{equation}
{\displaystyle P_{Flexible}\left(R,n\right)=\frac{2\left(2\pi\right)^{-2}}{R}\int_{0}^{\infty}dk\frac{\sin\left(kR\right)\left[\sin k\right]^{n}}{k^{n-1}},}\label{eq:flex}\end{equation}
which is the exact expression for the polymer propagator of the Random
Flight Model\cite{Freed}. Similarly, in the limit of $\alpha\rightarrow\infty$,
we can perform the following expansion valid for large values of $\alpha$

\[
{\displaystyle \frac{\left[\sinh\left\{ \sqrt{\alpha^{2}-k^{2}\nu_{\alpha,n}^{2}}\right\} \right]^{n-1}}{\left[\sqrt{\alpha^{2}-k^{2}\nu_{\alpha,n}^{2}}\right]^{n-1}}\simeq\frac{\exp\alpha\left(n-1\right)\exp\left[-\left(n-1\right)k^{2}\nu_{\alpha,n}^{2}/\alpha\right]}{\left[2\alpha\right]^{n-1}},}\]
and compute the polymer propagator. The result is 

\begin{equation}
{\displaystyle P_{rigid}\left(R,n,\alpha\right)\simeq\frac{2\left(2\pi\right)^{-2}}{4Rn}\frac{2\pi\exp\alpha\left(n-1\right)}{\left[2\alpha\right]^{n-1}}\delta\left(R-n\right),}\label{eq:rigido}\end{equation}
which is the polymer propagator of an infinitely stiff polymer chain\cite{Gennes}.
This propagator is not normalized.

We now combine the approximate expression of the characteristic function,
Eq. (\ref{eq:chacfinal}), with the Generalized Borel Transform to
compute the polymer propagator and the single chain structure factor
of the model.

We start the evaluation of the polymer propagator by rewriting Eq.
(\ref{eq:dis}) as follows

\begin{equation}
{\displaystyle P\left(R,n,\alpha\right)=\frac{{\displaystyle J\left(R-1-\left(n-1\right)g_{\alpha,n},n,\alpha\right)-J\left(R+1+\left(n-1\right)g_{\alpha,n},n,\alpha\right)}}{{\displaystyle 4\pi^{2}R\left[1+\left(n-1\right)g_{\alpha,n}\right]}},}\label{eq:propa}\end{equation}
where $J\left(r,n,\alpha\right)$ is defined by the mathematical expression 

\begin{equation}
{\displaystyle {\displaystyle J\left(r,n,\alpha\right)\equiv\int_{0}^{\infty}dk\left[\cos\left[kr\right]\left(\frac{{\displaystyle \sinh\left(\sqrt{\alpha^{2}-w^{2}\nu_{\alpha,n}^{2}}\right)}}{{\displaystyle \sqrt{\alpha^{2}-k^{2}\nu_{\alpha,n}^{2}}}}\right)^{n-1}\right].}}\label{eq:pimf}\end{equation}
This integral is evaluated exactly using GBT\cite{tecnica2,tecnica3,tecnica1}. A brief summary of the GBT technique can be found in Appendix A.

We present the most important steps of the calculation hereafter and
leave all the mathematical details for Appendix B. We first define
an auxiliary function $G\left(b,n,\alpha\right)$ as follows

\begin{equation}
{\displaystyle G\left(b,n,\alpha\right)\equiv\int_{0}^{\infty}dw\left[\exp\left(-wb\right)\left(\frac{{\displaystyle \sinh\left(\sqrt{\alpha^{2}-w^{2}\nu_{\alpha,n}^{2}}\right)}}{{\displaystyle \sqrt{\alpha^{2}-w^{2}\nu_{\alpha,n}^{2}}}}\right)^{n-1}\right],\quad b\geq0}\label{eq:gb}\end{equation}
from which the function $J\left(r,n,\alpha\right)$ can be computed
as the analytic continuation to the complex plane 

\begin{equation}
J\left(r,n,\alpha\right)=Re\left\{ G\left(b=-ir,n,\alpha\right)\right\} .\label{eq:disg}\end{equation}
Consequently, $J\left(r,n,\alpha\right)$ can be evaluated from the
Laplace Transform given by Eq. (\ref{eq:gb}). Following the technique
we can write 

\begin{equation}
{\displaystyle G\left(b,n,\alpha\right)=\begin{array}{c}
\lim\\
N\rightarrow\infty\end{array}\left(-\right){\displaystyle ^{N}\begin{array}{c}
\underbrace{\int db\cdots\int db}\\
N\end{array}}\frac{\Gamma\left(N+1\right)}{b^{N+1}}\left(\frac{{\displaystyle \sinh\sqrt{\alpha^{2}-\left(\frac{N\nu_{\alpha,n}}{b}\right)^{2}}}}{{\displaystyle \sqrt{\alpha^{2}-\left(\frac{N\nu_{\alpha,n}}{b}\right)^{2}}}}\right)^{n-1}.}\label{eq:gelimite}\end{equation}
The $N$ integrations are computed using well-established properties
and expansions of the functions $\sin\left(x\right)$ and $\left[1-x^{2}\right]^{-\mu}$\cite{tablarusa}.
After some straightforward algebra, we can write the analytical solution
of $G\left(b,n,\alpha\right)$ for any even number of segments as
follows 

\begin{equation}
\begin{array}{c}
{\displaystyle {\displaystyle G\left(b,n,\alpha\right)=\sum_{k=0}^{\frac{n-2}{2}}\left(-\right)^{k}\left(\begin{array}{c}
n-1\\
k\end{array}\right)\sum_{\beta=0}^{\infty}\alpha^{2\beta}}}\\
{\displaystyle {\displaystyle \times\frac{{\displaystyle _{3}F_{2}\left(\left[1,\frac{1}{2},\beta+1\right],\left[\beta+\frac{n+1}{2},\beta+\frac{n}{2}\right],-\frac{\left[\left(n-2k-1\right)\nu_{\alpha,n}\right]^{2}}{b^{2}}\right)}}{{\displaystyle \left(n-2k-1\right)^{-n-2\beta+1}2^{n-2}\Gamma\left(2\beta+n\right)b}}}},\end{array}\label{eq:gbfinal}\end{equation}
where $_{3}F_{2}\left(\left[,,\right],\left[,\right],x\right)$ and
$\Gamma\left(x\right)$ are the Generalized Hypergeometric\cite{hyper}
and Gamma\cite{abramovich} functions, respectively .

Replacing Eq.(\ref{eq:gbfinal}) into Eq.(\ref{eq:disg}) and computing
the analytic continuation to the complex plane through the substitution
$b=-ir$ we obtain 

\begin{equation}
\begin{array}{c}
{\displaystyle J\left(r,n,\alpha\right)=\frac{4}{2^{n}r}\sum_{k=0}^{\frac{n-2}{2}}\left(-\right)^{k+1}\left(\begin{array}{c}
n-1\\
k\end{array}\right)\sum_{\beta=0}^{\infty}\frac{\left(n-2k-1\right)^{2\beta+n-1}\alpha^{2\beta}}{\Gamma\left(2\beta+n\right)}}\\
{\displaystyle \times Im\left\{ _{3}F_{2}\left(\left[1,\frac{1}{2},\beta+1\right],\left[\beta+\frac{n+1}{2},\beta+\frac{n}{2}\right],\frac{\left[\left(n-2k-1\right)\nu_{\alpha,n}\right]^{2}}{r^{2}}\right)\right\} }.\end{array}\label{eq:disim}\end{equation}
The imaginary part of the Generalized Hypergeometric function is calculated
using its well-known analytical properties\cite{hyper}. This function
is an analytic function for values of the modulus of the argument
$z$ less than one and its continuation to the rest of complex plane
generates a cut on the positive real axis starting at $Re\left(z\right)=1$.
This implies that only values of the argument, $\frac{\left[\left(n-2k-1\right)\nu_{\alpha,n}\right]^{2}}{r^{2}}$,
larger or equal to one will contribute to the imaginary part of $_{3}F_{2}\left(z\right)$.
Consequently, this condition reduces the number of terms in the $k-$sum
such that the last term of Eq.(\ref{eq:gbfinal}) is $k=\left[\frac{n-1-r/\nu_{\alpha,n}}{2}\right]$.
The explicit evaluation of $Im\left\{ _{3}F_{2}\left(z\right)\right\} $
can be found in Appendix C. 

Finally, we replace Eq.(\ref{eq:hyperim}) into Eq.(\ref{eq:disim})
to obtain the exact expression for $J\left(r,n,\alpha\right)$ 

\begin{equation}
\begin{array}{c}
{\displaystyle J\left(r,n,\alpha\right)=\frac{\pi}{2\nu_{\alpha,n}}\sum_{k=0}^{\left[\frac{n-1-r/\nu_{\alpha,n}}{2}\right]}\left(-\right)^{k}\left(\begin{array}{c}
n-1\\
k\end{array}\right)\sum_{\beta=0}^{\infty}\frac{\left(\alpha\right)^{2\beta}}{\left[\beta!\right]^{2}}\sum_{L=0}^{\beta}\left(\begin{array}{c}
\beta\\
L\end{array}\right)}\\[0.2in]
{\displaystyle \times\left(\frac{r}{\nu_{\alpha,n}}\right)^{L}\left(\frac{n}{2}-k-\frac{r}{2\nu_{\alpha,n}}-\frac{1}{2}\right)^{2\beta+n-2-L}\frac{{\displaystyle \left(2\beta-L\right)!}}{{\displaystyle \left(2\beta+n-2-L\right)!}}}.\end{array}\label{eq:disfin}\end{equation}
Note that Eq. (\ref{eq:disfin}) is just a sum of polynomials in $r$.
The sum over the index $k$ is the one obtained for the Random Flight
Model\cite{RFM} and imposes the finite extensibility of the polymer
chain. The sums over the indexes $\beta$ and $L$ are a consequence
of the stiffness of the polymer backbone. The expression of $J\left(r,n,\alpha\right)$ can be rewritten as follows 

\begin{equation}
\begin{array}{c}
{\displaystyle J\left(r,n,\alpha\right)=\frac{\pi}{2\nu_{\alpha,n}}\sum_{k=0}^{\left[\frac{n-1-r/\nu_{\alpha,n}}{2}\right]}\frac{\left(-\right)^{k}}{\left(n-3\right)!}\left(\begin{array}{c}
n-1\\
k\end{array}\right)\left(\frac{n}{2}-k-\frac{r}{2\nu_{\alpha,n}}-\frac{1}{2}\right)^{n-2}}\\
{\displaystyle \times\int_{0}^{1}dz\,\left(1-z\right)^{n-3}\, I_{0}\left(2\alpha\left(\frac{n}{2}-k-\frac{r}{2\nu_{\alpha,n}}-\frac{1}{2}\right)\sqrt{z^{2}+z\frac{\frac{r}{\nu_{\alpha,n}}}{\left(\frac{n}{2}-k-\frac{r}{2\nu_{\alpha,n}}-\frac{1}{2}\right)}}\right),}\end{array}\label{eq:disfin2}\end{equation}
 which can be used for further approximation if so desired. 

The expression for $J\left(r,n,\alpha\right)$ given by Eq. (\ref{eq:disfin})
was derived for even number of segments but, its validity for odd
number of segments larger than two can be proved by analytic continuation.
Finally, in order to obtain the polymer propagator for semiflexible
chains we have to replace Eq. (\ref{eq:disfin}) into Eq. (\ref{eq:propa}). Observe that, after the replacement, Eq. (\ref{eq:propa}) recovers the known exact solution of the Random Flight Model\cite{RFM} when the limit $\alpha \rightarrow 0$ is taken. Indeed, the only term different from zero is the one for which $\beta = 0$.

Finally, we conclude our calculations of the WCM by computing the single chain structure factor which is defined
by the following formula \cite{Benoit,Mattice}

\begin{equation}
{\displaystyle S\left(k,n,\alpha\right)=\frac{1}{n+1}+\frac{2}{\left(n+1\right)^{2}}\sum_{ni=1}^{n}\frac{\left(n-ni+1\right)}{Z_{ni}\left(\alpha\right)}K\left(k,\alpha,ni\right).}\label{eq:structuralfactor}\end{equation}

\section{Results and discussion}

Figures 1, 2 and 3 show the prediction of the normalized polymer propagator,
Eq. (\ref{eq:propa}), as function of the end-to-end distance for
chains with 5, 10 and 30 Kuhn segments, and different values of the semiflexibility
parameter $\alpha$. The numerical evaluation of the propagator was
done using Eq. (\ref{eq:disfin}). The sum over the index $\beta$
converges quickly even for large values of the semiflexibility parameter
$\alpha$. Indeed, even with less than 80 terms in the sum we obtained
a relative precision of $10^{-4}$. The figures clearly show that
the location of the peak in the polymer propagator (multiplied by
$R^{2}$) moves toward larger values of $R$ when the stiffness of
the polymer backbone increases. This behavior is in good qualitative
agreement with previous results arising from computer simulation studies\cite{frey}
and theoretical approaches based on the Maximum Entropy Principle\cite{Winkler}.
This is the correct result because the stiffer the polymer backbone,
the higher the energetic penalty to bend the chain. Consequently,
those configurations of the macromolecule with small end-to-end distance
will be more and more hindered as the stiffness increases while those
configurations with large end-to-end distance should be more and more
favored. Therefore, the peak should shift toward larger values of
$R$ when the stiffness increases. 

Figure 4 shows the polymer propagator for polymer chains with 5, 10
and 30 Kuhn segments and a fixed value of the semiflexibility parameter
$\alpha(=3.0)$. This figure shows that the longer the polymer is,
the more it behaves like a flexible chain since the location of peak
(=end-to-end distance divided by the contour length) moves toward
smaller values. In other words, the longer the polymer is, the less
relevant the stiffness of the backbone becomes. 

Figures 5, 6 and 7 show the behavior of $S\left(k,n,\alpha\right)$, Eq. (\ref{eq:structuralfactor}),
as a function of $k$ for different values of $\alpha$ and three
values of $n$ (5, 10 and 30). The figures clearly show a decrease
of the single chain structure factor with increasing $k$ until it
reaches a plateau at infinite $k$. Note that our computations predict
that the decrease of the single chain structure factor for small values
of $k$ should be faster in the case of stiff polymers than in the
case of flexible ones. This is a consequence of the fact that rigid
polymers have a larger radius of gyration than flexible ones for a
fixed chain length. In addition, the decrease of the structure factor
for large values of $k$ is faster for flexible polymers than for
stiff ones. Indeed, our computations predict that the structure factor
goes as $k^{-\theta}$ for large values of $k$ just before the plateau
is reached. For values of $n$=5, 10 and 30, the values of $\theta$
that we got were 1, 1.08 and 1.3 for $\alpha$=10 (rigid) and 1.64,
1.9 and 2 for $\alpha$=0.33 (flexible). These results are in good
agreement with the fact that the structure factor of polymers with
large chain length should scale as $k^{-d}$ for large $k$ values
where $d$ is the fractal dimension of the object (2 for a flexible
polymer and 1 for a rigid polymer). Consequently, these results imply
that for short polymer chains ($n=5$), a value of $\alpha=10$ is
high enough to make this polymer behave like stiff rod, $k^{-1}$.
On the other hand, as the chain length increases we observe that $\alpha=10$
is not high enough to make the polymer behave as a rigid rod and deviations
from the power law $k^{-1}$ are observed. In the case of $\alpha=0.33$
our calculations predict that a value of $n$ equal to five is not
high enough to recover the scaling behavior of flexible chains, $k^{-2}$.
But, as the number of segments increases, the exponent approaches
the value of two, indicating that the polymer chain behaves more and
more like a flexible one. This result was also showed in Figure 4.
The figures also show that the approximation we developed in the previous
section gives a smooth crossover behavior from the rigid to the flexible
limit.

In order to make the presentation of our work more balanced and objective, we proceed to compare our results with the predictions of two other models. We start with our prediction for the single chain structure factor, $S \left( k \right)$, and compare it with the expression obtained by Kholodenko\cite{Kholodenko}. It has been shown that Kholodenko's result can describe experimental data quantitatively\cite{Ballauf}. Thus, a comparison between our expression for the single chain structure factor and Kholodenko's will help us gauge the quality of the approximations used in our treatment of the WCM. Figures 8 and 9 show the comparison for polymers with $n=30$ and two different values of the semiflexibility parameter (shown in the plots). We have checked that other values of the parameters $\alpha$ or $a$ and $n$ give quantitative agreements of similar quality. In all the cases studied we found that the relationship $\alpha=2 a$ always gives excellent quantitative agreement between the predictions of both models. Thus, our single chain structure factor should agree very well with the experimental data of Ballauff and coworkers. Still, since both models have different origins, some very small differences can be observed in the case of flexible chains (Figure 8). 

We now proceed to compare our expression of the polymer propagator with the one computed by Wilhelm and Frey\cite{Frey}. Figure 10 shows this comparison for the case of polymers with $n=5$. The continuous curves are the results of our calculation with $\alpha=1, 3 , 5$ and $10$. The dashed curves were constructed based on the work of Wilhelm and Frey where we adjusted the bending modulus such that the location of the peak in the propagator matched our results. This gives a better picture of the differences and similarities between both results. Figure 10 shows quantitative agreement between both results when the stiffness is low. As the stiffness increases Fig. 10 shows that the qualitative behavior of both propagators is still the same. For example, both results predict that the location of the peak moves toward larger values of the end-to-end distance and that the distribution becomes narrower. The main difference between both results is quantitative in nature. Our distribution becomes narrower than the one predicted by Wilhelm and Frey's work by a factor of two approximately which, in turn, generates a higher peak (the distributions are normalized). 

Finally, let us conclude this section by rationalizing the origin of the discrepancy between the propagators and the reason why this does not affect the structure factor significantly. We first note that differences in the propagators are to be expected because both, ours and Frey's, calculations are based on different approximations and versions of the model (continuum or discrete). Let us now proceed to rationalize the origin of the discrepancy between both predictions. Equations (\ref{eq:distri}) and (\ref{eq:structuralfactor}) define the polymer propagator and single chain structure factor in terms of the characteristic function. Observe that the propagator is a Fourier transform of the characteristic function. Consequently, the oscillatory nature of the complex exponential generates partial cancellation of the contributions to the integral arising from different parts of the interval of integration. This cancellation magnifies any inaccuracies made in the approximation of the characteristic function. Moreover, the consequences of this cancellation become more and more important as the polymer becomes stiffer because the characteristic function itself adopts an oscillatory behavior which, in principle, is out of phase with respect to the complex exponential. Therefore, the stiffer the polymer is, the more important the consequences of the approximation become. On the other hand, the evaluation of the structure factor from the characteristic function does not involve any oscillatory function. Consequently, any small inaccuracy made in the approximation of the characteristic function will remain small in the expression of the structure factor, as shown previously.

\section{conclusions}

The results obtained in this paper show that the Generalized Borel Transform is a very
useful computational tool for the statistical mechanics of single semiflexible
polymer chains. Indeed, the results presented in this paper clearly show that GBT
is able to compute polymer propagators for single chain problems exactly.
This capability of the technique is a direct consequence of its mathematical
simplicity (the GBT requires elements of basic calculus and some fundamental
knowledge of complex variables). Consequently, it does not add any
mathematical complexity to the physics of the starting model. 

Our analysis of the Wormlike Chain Model was based on an approximate
expression of the characteristic function. The exact evaluation of
this function is not possible at present. Therefore, we developed
a new approximation that preserves the most relevant physical characteristics
of the model intact. Specifically, our approach keeps the thermodynamics
of the model, the flexible and rigid limits, the mean square end-to-end
distance and the finite extensibility of the model intact while providing
an approximate expression for the characteristic function for intermediate
values of the stiffness of the polymer chain. The polymer propagator
was obtained exactly from the approximate characteristic function
using GBT. Note that the propagator is approximate not because of
the GBT, which computes this quantity exactly, but because of the
approximate nature of the characteristic function. Our expression
for the propagator shows a peak that shifts toward larger values of
the end-to-end distance as the stiffness of the polymer backbone is
increased, in agreement with other theoretical and computational treatments
of the model. We also found that, in the low wave vector region, the
structure factor decreases faster with increasing wave vector when
the stiffness increases. This was rationalized in terms of the behavior
of the radius of gyration. Similarly, we found that, in the large
wave vector region, the structure factor of flexible chains decreases
faster than the one of rigid polymers. This was compared with the
behavior of very long polymer chains whose behavior for large wave
vectors is known exactly. We also compared of the predictions of our calculation with established results for the single chain structure factor and polymer propagator. Excellent quantitative agreement was observed between our prediction for the single chain structure factor and the one predicted by Kholodenko's model. The polymer propagator was compared with the prediction of Wilhelm and Frey. Very good quantitative agreement was observed between the predictions of both models for low values of the stiffness. For stiff polymers, quantitative deviations were observed and the origin of the deviations was rationalized.  

The proposed approach to semiflexible polymers can also address semiflexible
polymers with other topologies like ring and m-arm star polymers.
The procedure should be similar to the one presented in this paper
but, the characteristic function will have a different mathematical
expression.

We conclude this section with a discussion of the characteristic function
of the WCM which limits our ability to solve this model exactly. Let
us rewrite this function. The expression is

\begin{equation}
{\displaystyle K\left(\left\{ \mathbf{k}_{j}\right\} ,\alpha,n,l\right)\equiv\int d\left\{ \mathbf{R}_{k}\right\} \prod_{j=1}^{n}\delta\left(\left|\mathbf{R}_{j}\right|-l\right)\exp\left(i\sum_{j=1}^{n}\mathbf{R}_{j}\cdot\mathbf{k}_{j}+\alpha\sum_{i=1}^{n-1}\mathbf{R}_{i+1}\cdot\mathbf{R}_{i}\right)},\end{equation}
where we have replaced the wave vector $\mathbf{k}$ by a group of
wave vectors $\mathbf{k}_{j}$. This expression is the one of the
characteristic function when all the $\mathbf{k}_{j}$ are equal to
$\mathbf{k}$. Observe that if we replace $\mathbf{k}_{j}$ by $i\mathbf{b}_{j}$
where $i$ is the imaginary unit, then 

\begin{equation}
{\displaystyle K\left(\left\{ \mathbf{b}_{j}\right\} ,\alpha,n,l\right)\equiv\int d\left\{ \mathbf{R}_{k}\right\} \prod_{j=1}^{n}\delta\left(\left|\mathbf{R}_{j}\right|-l\right)\exp\left(-\sum_{j=1}^{n}\mathbf{R}_{j}\cdot\mathbf{b}_{j}+\alpha\sum_{i=1}^{n-1}\mathbf{R}_{i+1}\cdot\mathbf{R}_{i}\right)},\end{equation}
which has the form of a Laplace Transform in $3n$ dimensions. But,
the GBT computes Laplace Transforms very accurately or even exactly.
Consequently, a generalization of GBT to many dimensions might lead
to an exact or very accurate expression of the characteristic function
of the WCM. This expression can be further used to compute the polymer
propagator using GBT. Thus, we speculate that such extension of the
GBT technique might allow us to solve very accurately or even exactly
the Wormlike Chain Model. In general, such extension of GBT might
allow us to solve other models of polymer chains of the form

\begin{equation}
{\displaystyle K\left(\left\{ \mathbf{k}_{j}\right\} ,n,\mathrm{parameters}\right)\equiv\int d\left\{ \mathbf{R}_{k}\right\} \exp\left(i\sum_{j=1}^{n}\mathbf{R}_{j}\cdot\mathbf{k}_{j}+\mathrm{H}\left[\left\{ \mathbf{R}_{l}\right\} ,\mathrm{parameters}\right]\right)},\end{equation}
where $\mathrm{H}\left[\left\{ \mathbf{R}_{l}\right\} ,...\right]$
is the Hamiltonian of the model. Thus, helical wormlike polymers and
other models might be mathematically tractable with this generalization
of the GBT.

\section{Acknowledgments}

We acknowledge the National Science Foundation, Grant \# CHE-0132278
(CAREER), the Ohio Board of Regents Action Fund, Proposal \# R566
and The University of Akron for financial support.

\appendix

\section{THE Generalized Borel Transform (GBT)}

Let us briefly present the mathematical aspects of the GBT in connection
with the computation of the Laplace-Mellin transform\cite{tecnica2,tecnica3,tecnica1}.
We start with the expression

\begin{equation}
S\left(g,a,n\right)=\int_{0}^{\infty}x^{n}H\left(x,a\right)\exp\left(-gx\right)dx,\qquad g>0\label{int}\end{equation}
 where we have explicitly extracted a factor $x^{n}$ from the function
to be transformed.

Defining the Generalized Borel Transform (GBT) of $S$ as

\begin{equation}
B_{\lambda}\left(s,a,n\right)\equiv\int_{0}^{\infty}\exp\left[s/\eta\right]\left[\frac{1}{\lambda\eta}+1\right]^{-\lambda s}S\left(g,a,n\right)d\left(1/\eta\right),\quad Re\left(s\right)<0\label{borel}\end{equation}
 where $\lambda$ is any real positive non-zero value and $1/\eta\equiv\lambda\left(\exp\left(g/\lambda\right)-1\right)$,
we can invert Eq. (\ref{borel}) in the following way

\begin{equation}
S\left(g,a,n\right)=2\lambda^{2}\left(1-\exp\left(-g/\lambda\right)\right)\int_{-\infty}^{\infty}\int_{-\infty}^{\infty}\exp\left[G\left(w,t,g,\lambda,a,n\right)\right]dwdt,\label{doble}\end{equation}
 where the explicit expression of $G\left(w,t,g,\lambda,a,n\right)$
is not important for our present purposes (for more details see Ref.
\cite{tecnica2}).

The expression given by Eq. (\ref{doble}) is valid for any non-zero,
real and positive value of the parameter $\lambda$. But, the resulting
expression for $S\left(g,a,n\right)$ does not depend on $\lambda$
explicitly. Thus, we can choose the value of this parameter in such
a way that it allows us to solve Eq. (\ref{doble}). The dominant
contribution to the double integral is obtained using steepest descent
\cite{norman,copson} in the combined variables $\left[t,w\right]$.
In doing so, one first computes the saddle point $t_{o}\left(g,a,n\right)$
and $w_{o}\left(g,a,n\right)$ in the limit $\lambda\gg1$ and then
checks the positivity condition \cite{jefrey} (the Hessian of $G$
at this point should be positive) obtaining \begin{equation}
t_{o}=\ln\left[\frac{x_{o}^{2}\left(g,a,n\right)}{f\left(x_{o}\left(g,a,n\right),a,n\right)}\right]=t_{o}\left(g,a,n\right)\quad,\quad w_{o}=\ln\left[x_{o}\left(g,a,n\right)\right]=w_{o}\left(g,a,n\right),\label{tw}\end{equation}
 where $x_{o}\left(g,a,n\right)$ is the real and positive solution
of the implicit equation coming from the extremes of the function
$G$ in the asymptotic limit in $\lambda$. Therefore, one obtains
the following equation

\begin{equation}
x_{o}^{2}g^{2}=f\left(x_{o},a,n\right)\left[f\left(x_{o},a,n\right)+1\right],\label{gx}\end{equation}
 where

\begin{equation}
f\left(x_{o},a,n\right)\equiv1+n+x_{o}\frac{d\ln\left[H\left(x_{o},a\right)\right]}{dx_{o}}.\label{fx}\end{equation}

In the range of the parameters where $f\left(x_{o},a,n\right)\gg1$
which is fulfilled when $n\gg1$, and assuming that there is only
one saddle point, we can retain the first order in the expansion of
$G$ around the saddle point. 

Finally, we obtain the approximate expression for the starting function
$S\left(g,a,n\right)$ \begin{equation}
S_{Aprox}\left(g,a,n\right)=\sqrt{2\pi}e^{-1/2}\frac{\sqrt{f\left[x_{o},a,n\right]+1}}{\sqrt{D\left[x_{o},a,n\right]}}\left[x_{o}\right]^{n+1}H\left[x_{o},a\right]\exp\left[-f\left[x_{o},a,n\right]\right],\label{eq:res}\end{equation}
where

\begin{equation}
D\left(x_{o},a,n\right)=-x_{o}\,\frac{df\left(x_{o},a,n\right)}{dx_{o}}\left[1/2+f\left(x_{o},a,n\right)\right]+f\left(x_{o},a,n\right)\left[1+f\left(x_{o},a,n\right)\right].\label{dx}\end{equation}

Note that the expression given by Eq. (\ref{eq:res}) is valid for
functions $H\left(x,a\right)$ that fulfill the following general
conditions:

1) the relation given by Eq. (\ref{gx}) must be biunivocal.

2) $D\left(x_{o},a,n\right)$ has to be positive and $\left[x_{o}\frac{df\left(x_{o},a,n\right)}{dx_{o}}-2f\left(x_{o},a,n\right)\right]$
has to be negative in $x_{o}$.

3) $f\left(x_{o},a,n\right)\gg1$. In particular, this condition is
fulfilled when $n\gg1.$

These conditions provide the range of values of the parameters where
the approximate solution, Eq. (\ref{eq:res}), is valid. 

In summary, the GBT provides an approximate solution ,Eq. (\ref{eq:res}),
to amplitudes with the mathematical form given by Eq. (\ref{int}).
The calculation consists of solving the implicit equation Eq. (\ref{gx})
for $n\gg1$ to obtain the saddle point and replace it into Eq. (\ref{eq:res}).

Let us now focus on amplitudes with the mathematical form of a Laplace
transform

\begin{equation}
S\left(g,a\right)=\int_{0}^{\infty}H\left(x,a\right)\exp\left(-gx\right)dx\qquad g>0.\label{eq:fi}\end{equation}
 This kind of amplitudes can be mapped onto expressions of the form
given by Eq. (\ref{int}). In order to use the GBT on Eq. (\ref{eq:fi}),
we use the following relationship between Eq. (\ref{eq:fi}) and Eq.
(\ref{int})

\begin{equation}
S\left(g,a,n\right)=\left(-\right)^{n}\frac{\partial^{n}}{\partial g^{n}}S\left(g,a\right),\label{eq:19}\end{equation}
 which can be inverted to give

\begin{equation}
S\left(g,a\right)=\left(-\right)^{n}\underbrace{\int dg\cdots\int dg}_{n}S\left(g,a,n\right)+\sum_{p=0}^{n-1}c_{p}\left(a,n\right)g^{p}.\label{fiapro}\end{equation}
 The finite sum comes from the indefinite integrations. Note that
all the coefficients vanish whenever the Laplace transform, Eq. (\ref{eq:fi}),
fulfills the following asymptotic behavior

\begin{equation}
\lim_{g\rightarrow\infty}S\left(g,a\right)=0.\label{eq:asymp}\end{equation}
 In addition, the expression, given by Eq. (\ref{fiapro}), is valid
for any value of $n,$ in particular for $n\gg1.$ Consequently, if
Eq. (\ref{eq:asymp}), is fulfilled, then the analytical solution
reads

\begin{equation}
S\left(g,a\right)=\begin{array}{c}
\lim\\
n\rightarrow\infty\end{array}\left(-\right)^{n}\underbrace{\int dg\cdots\int dg}_{n}S_{Aprox}\left(g,a,n\right),\label{fff}\end{equation}
 where, for $n\gg1$, we can use the expression given by Eq. (\ref{eq:res}).
It is important to note that it is the limit $n\rightarrow\infty$
that makes the saddle point solution, Eq. (\ref{eq:res}), an exact
solution for Eq. (\ref{doble}). Thus, as long as the $n$ indefinite
integrals can be done without approximations, as it is in our case,
the result for $S$ is exact.

\section{evaluation of the polymer propagator}

We start the evaluation of the polymer propagator by rewriting Eq.
(\ref{eq:gb}) as follows 

\begin{equation}
\begin{array}{c}
{\displaystyle G\left(b,n,\alpha\right)\equiv\frac{\partial^{n-1}}{\partial c^{n-1}}\left\{ \int_{0}^{\infty}dw\left[\exp\left(-wb\right)\exp\left(c\frac{\sinh\left\{ \sqrt{\alpha^{2}-w^{2}\nu_{\alpha,n}^{2}}\right\} }{\sqrt{\alpha^{2}-w^{2}\nu_{\alpha,n}^{2}}}\right)\right]\right\} _{c=0}}\\
{\displaystyle =\frac{\partial^{n-1}}{\partial c^{n-1}}\left\{ GA\left(b,\alpha,c\right)\right\} _{c=0}},\end{array}\label{eq:gb2}\end{equation}
where $GA\left(b,\alpha,c\right)$ is 

\begin{equation}
{\displaystyle GA\left(b,\alpha,c\right)\equiv\int_{0}^{\infty}dw\left[\exp\left(-wb\right)H\left(w,\alpha,c\right)\right]},\label{eq:ga}\end{equation}
and $H\left(w,\alpha,c\right)$ is given by the expression

\begin{equation}
H\left(w,\alpha,c\right)\equiv\exp\left(c\frac{\sinh\left\{ \sqrt{\alpha^{2}-w^{2}\nu_{\alpha,n}^{2}}\right\} }{\sqrt{\alpha^{2}-w^{2}\nu_{\alpha,n}^{2}}}\right).\label{eq:dfrt}\end{equation}

The integral expressed by Eq.(\ref{eq:ga}) satisfies all the requirements
of the GBT technique. Then, we evaluate it in the following way 

\begin{equation}
{\displaystyle GA\left(b,\alpha,c\right)=\begin{array}{c}
\lim\\
N\rightarrow\infty\end{array}\begin{array}{c}
\left(-\right)^{N}\underbrace{\int db\cdots\cdots\cdots\int db}GA_{N},\\
N\end{array}}\label{eq:gaprox}\end{equation}
where 

\begin{equation}
{\displaystyle GA_{N}\left(b,\alpha,c\right)\equiv\int_{0}^{\infty}dw\left[w^{N}\exp\left(-wb\right)H\left(w,\alpha,c\right)\right]}.\label{eq:gaproxeneuno}\end{equation}

In the asymptotic limit of $N\rightarrow\infty$ the GBT provides
an analytical solution for Eq.(\ref{eq:gaproxeneuno}). Following
the technique, we solve the implicit equation, Eq. (\ref{gx}), for the saddle point $w_{o}$. The asymptotic solution
is

\begin{equation}
{\displaystyle w_{o}\simeq\frac{N+3/2}{b}\quad N\gg1.}\label{eq:sp}\end{equation}
Replacing this expression for $w_{o}$ in the expression provided
by the GBT, Eq. (A7), we obtain

\begin{equation}
{\displaystyle GA_{N}\left(b,\alpha,c\right)\simeq\frac{\Gamma\left(N+1\right)}{b^{N+1}}H\left(\frac{N+3/2}{b},\alpha,c\right)\quad N\gg1.}\label{eq:gan}\end{equation}

Furthermore, we replace Eq.(\ref{eq:gan}) into Eq.(\ref{eq:gaprox})
and the resulting expression into Eq.(\ref{eq:gb2}), and we exchange
the order of the operators in the resulting expression. In other words,
we first evaluate the $n$ derivatives with respect to $c$ and, afterward,
we take the limit of $c\rightarrow0$ to obtain the expression given
by Eq. (\ref{eq:gelimite}). Next, we solve the $N$ integrations
using standard properties and expansions of the functions $\sin\left(x\right)$
and $\left[1-x^{2}\right]^{-\mu}$\cite{tablarusa}, and write $G\left(b,n,\alpha\right)$
as follows 

\begin{equation}
G\left(b,n,\alpha\right)=\frac{1}{2^{n-2}}\sum_{k=0}^{\frac{n-2}{2}}\left(-\right)^{\frac{n-2}{2}+k}\left(\begin{array}{c}
n-1\\
k\end{array}\right)M\left(N,n,k,\alpha,b\right),\label{eq:gb3}\end{equation}
where $n$ is even and 

\begin{equation}
\begin{array}{c}
{\displaystyle M\left(n,k,\alpha,b\right)\equiv\begin{array}{c}
\lim\\
N\rightarrow\infty\end{array}\left(-\right)^{N}Im\sum_{r=0}^{\infty}\frac{\left(i\left(n-1-2k\right)\right)^{r}\left[N\nu_{\alpha,n}\right]^{r-n+1}}{r!}}\\
{\displaystyle \times\sum_{\beta=0}^{\infty}L_{\beta}^{-\frac{n-r-1}{2}-\beta}\left(0\right)\left(-\frac{\alpha^{2}}{N^{2}}\right)^{\beta}\Gamma\left(N+1\right)\int db\cdots\int db\frac{1}{b^{2+N-n+r-2\beta}},}\end{array}\label{eq:eme}\end{equation}
where $L_{\beta}^{\gamma}\left(x\right)$ are the Laguerre polynomials
\cite{tablarusa}.

Note that the only powers on $b$ in Eq. (\ref{eq:eme}) that fulfill
the asymptotic behavior of the function $G\left(b,n,\alpha\right)$
are those for which the condition $r\geq\left(n+2\beta-1\right)$
is satisfied. Therefore, the $N$ indefinite integrations are exactly
doable. The result is

\begin{equation}
{\displaystyle \int db\cdots\int db\frac{1}{b^{2+N-n+r-2\beta}}=\frac{\Gamma\left(2+r-n-2\beta\right)}{\Gamma\left(2+N-n+r-2\beta\right)}\frac{\left(-\right)^{N}}{b^{2+r-n-2\beta}}.}\label{eq:nint}\end{equation}
Replacing Eq. (\ref{eq:nint}) into Eq. (\ref{eq:eme}) and after
the change of variables $r=x+2\beta+n-1,$ we can write 

\begin{equation}
\begin{array}{c}
{\displaystyle M\left(n,k,\alpha,b\right)\equiv Im\frac{1}{b}\sum_{\beta=0}^{\infty}\left(i\left(n-2k-1\right)\right)^{2\beta+n-1}\left(-\alpha^{2}\right)^{\beta}\sum_{x=0}^{\infty}\left(\frac{i\left(n-2k-1\right)\nu_{\alpha,n}}{b}\right)^{x}}\\
{\displaystyle \times\frac{\Gamma\left(x+1\right)L_{\beta}^{\frac{x}{2}}\left(0\right)}{\Gamma\left(x+2\beta+n\right)}\begin{array}{c}
\lim\\
N\rightarrow\infty\end{array}\frac{N^{x}\Gamma\left(N+1\right)}{\Gamma\left(N+x+1\right)}}.\end{array}\label{eq:eme2}\end{equation}

Using the asymptotic properties of the Gamma function\cite{abramovich}

\begin{equation}
{\displaystyle \begin{array}{c}
\lim\\
N\rightarrow\infty\end{array}\frac{N^{x}\Gamma\left(N+1\right)}{\Gamma\left(N+x+1\right)}=1,}\label{eq:poiu}\end{equation}
 Eq. (\ref{eq:eme2}) finally reads

\begin{equation}
\begin{array}{c}
{\displaystyle M\left(n,k,\alpha,b\right)=\frac{1}{b}Im\sum_{\beta=0}^{\infty}\sum_{x=0}^{\infty}\left(i\left(n-2k-1\right)\right)^{2\beta+n-1}\frac{\left(-\alpha^{2}\right)^{\beta}}{\beta!}\left(\frac{i\left(n-2k-1\right)\nu_{\alpha,n}}{b}\right)^{x}}\\
{\displaystyle \frac{\Gamma\left(x+1\right)\Gamma\left(\frac{x+2}{2}+\beta\right)}{\Gamma\left(x+2\beta+n\right)\Gamma\left(\frac{x+2}{2}\right)}}\end{array}\label{eq:pepe}\end{equation}

The sum over $x$ is exactly doable. The result is 

\begin{equation}
M\left(n,k,\alpha,b\right)=\frac{1}{b}Im\sum_{\beta=0}^{\infty}\left(i\left(n-2k-1\right)\right)^{2\beta+n-1}\frac{\left(-\alpha^{2}\right)^{\beta}}{\beta!}FD\left(\beta,n,k,b\right),\label{eq:eme3}\end{equation}
where 

\begin{equation}
\begin{array}{c}
{\displaystyle FD\left(\beta,n,k,b\right)\equiv\frac{\Gamma\left(\beta+1\right)\,_{3}F_{2}\left(\left[1,\frac{1}{2},\beta+1\right],\left[\beta+\frac{n+1}{2},\beta+\frac{n}{2}\right],-\frac{\left[\left(n-2k-1\right)\nu_{\alpha,n}\right]^{2}}{b^{2}}\right)}{\Gamma\left(2\beta+n\right)}}\\
{\displaystyle +\frac{2i\left(n-2k-1\right)\nu_{\alpha,n}}{b}\frac{\Gamma\left(\beta+\frac{3}{2}\right)\,_{3}F_{2}\left(\left[\beta+\frac{3}{2},1,1\right],\left[\beta+\frac{n}{2}+1,\beta+\frac{n}{2}+\frac{1}{2}\right],-\frac{\left[\left(n-2k-1\right)\nu_{\alpha,n}\right]^{2}}{b^{2}}\right)}{\Gamma\left(2\beta+n+1\right)\sqrt{\pi}}.}\end{array}\label{eq:Fd}\end{equation}

Equation (\ref{eq:eme3}) clearly shows that the imaginary part affects
only the real part of the function $FD\left(\beta,n,k,b\right)$.
Thus, the final expression for $G\left(b,n,\alpha\right)$ is given
by Eq. (\ref{eq:gbfinal}).

\section{evaluation of the imaginary part of the generalized hypergeometric
function}

We evaluate $Im\left\{ _{3}F_{2}\left(z\right)\right\} $ using the
following integral representation of the Hypergeometric function\cite{tablarusa} 

\begin{equation}
\begin{array}{c}
{\displaystyle _{3}F_{2}\left(\left[1,\frac{1}{2},\beta+1\right],\left[\beta+\frac{n+1}{2},\beta+\frac{n}{2}\right],-\frac{\left(n-2k-1\right)^{2}}{b^{2}}\right)=\frac{b^{2\left(\beta+1\right)}}{B\left(1,2\beta+n-1\right)}}\\
{\displaystyle \times\frac{1}{\left(n-2k-1\right)^{2\beta+n-1}}\int_{0}^{n-2k-1}\left[n-2k-1-x\right]^{2\beta+n-2}\left[x^{2}+b^{2}\right]^{-\beta-1}dx},\end{array}\label{eq:otravez}\end{equation}
which is valid for values of $n$ equal or larger than two. The analytic
continuation to the complex plane is done as before through the replacement
$b=-ir$. Then, the imaginary part of the Hypergeometric function
is

\begin{equation}
\begin{array}{c}
\begin{array}{c}
{\displaystyle Im\left\{ _{3}F_{2}\left(\left[1,\frac{1}{2},\beta+1\right],\left[\beta+\frac{n+1}{2},\beta+\frac{n}{2}\right],\frac{\left(n-2k-1\right)^{2}}{r^{2}}\right)\right\} =}\\
{\displaystyle \frac{r^{2\left(\beta+1\right)}\left(-\right)^{\beta+1}}{B\left(1,2\beta+n-1\right)\left(n-2k-1\right)^{2\beta+n-1}}}\\
{\displaystyle \times Im\int_{0}^{n-2k-1}\left[n-2k-1-x\right]^{2\beta+n-2}\left[x^{2}-r^{2}\right]^{-\beta-1}dx}.\end{array}\end{array}\label{eq:imaginarypart}\end{equation}
Thus, we have to evaluate the expression

\begin{equation}
L\equiv Im\int_{0}^{n-2k-1}\left[n-2k-1-x\right]^{2\beta+n-2}\left[x-r\right]^{-\beta-1}\left[x+r\right]^{-\beta-1}dx.\label{eq:34er}\end{equation}

After analyzing the analytical behavior of the integrand we concluded
that we can exchange the operations of integration and imaginary part
to obtain

\begin{equation}
L=\int_{0}^{n-2k-1}\left[n-2k-1-x\right]^{2\beta+n-2}\left[x+r\right]^{-\beta-1}Im\left\{ \left[x-r\right]^{-\beta-1}\right\} dx.\label{eq:ele}\end{equation}
Thus, we have to compute 

\begin{equation}
LS=Im\left\{ \frac{1}{\left(x-r\right)^{\beta+1}}\right\} ,\label{eq:dfc3}\end{equation}
first and, afterward, we have to solve the integral given by Eq. (\ref{eq:ele}).

The analytical behavior of the function $\left(x-R\right)^{-\beta-1}$
is well known\cite{tablarusa}. It is an analytic function for $\left|x\right|>R$
but, its analytic continuation to the complex plane generates a cut
on the real axis in the range $-R<Re\left(x\right)<R$ which provides
its imaginary part. Writing

\begin{equation}
{\displaystyle \frac{1}{\left(x-R\right)^{\beta+1}}=\frac{1}{\left(x-R\right)^{\beta}}\frac{1}{\sqrt{x-R}\sqrt{x-R}}}\label{eq:qwa1}\end{equation}
for integer values of $\beta$ and $x>R$, and using the following
integral representation 

\begin{equation}
{\displaystyle \frac{1}{\sqrt{x-R}}=\int_{0}^{\infty}dy\exp\left[-y\sqrt{x-R}\right]},\label{eq:asw2}\end{equation}
valid for $x>R$, we obtain the result

\begin{equation}
{\displaystyle Im\left\{ \frac{1}{\left(x-R\right)^{\beta+1}}\right\} =\frac{\left(-\right)^{\beta+1}}{\left(R-x\right)^{\beta}}Re\left\{ \int_{0}^{\infty}du\exp\left[-iu\left(R-x\right)\right]\right\} ,}\label{eq:rangointe}\end{equation}
where $x<R$.

Using parity's properties of the function $\cos\left(\theta\right),$
we extend the range of the integration in Eq. (\ref{eq:rangointe})
to obtain 

\begin{equation}
{\displaystyle Im\left\{ \frac{1}{\left(x-R\right)^{\beta+1}}\right\} =\frac{\pi}{\left(x-R\right)^{\beta}}\delta\left(x-R\right)=\frac{\pi\left(-\right)^{\beta}}{\beta!}\frac{\partial^{\beta}}{\partial\left(x-R\right)^{\beta}}\delta\left(x-R\right).}\label{eq:delta2}\end{equation}

Finally, we replace the expression given by Eq. (\ref{eq:delta2})
into Eq. (\ref{eq:ele}) and perform the change of variables $y=x-r$
to obtain

\begin{equation}
{\displaystyle L=\frac{\pi\left(-\right)^{\beta}}{\beta!}\int_{-r}^{n-2k-1-r}F_{\beta k}\left(y,n,r\right)\frac{\partial^{\beta}}{\partial y^{\beta}}\delta\left(y\right)dy,}\label{eq:ieur}\end{equation}
where $F_{\beta k}\left(y,n,r\right)$ is defined as

\begin{equation}
F_{\beta k}\left(y,n,r\right)\equiv\left[n-2k-1-r-y\right]^{2\beta+n-2}\left[y+2r\right]^{-\beta-1}.\label{eq:efepoly}\end{equation}

Integrating by parts $\beta$ times, we obtain the following final
expression 

\begin{equation}
{\displaystyle L=\frac{\pi}{\beta!}\left\{ \frac{\partial^{\beta}F_{\beta k}\left(y,n,r\right)}{\partial y^{\beta}}\right\} _{y=0}.}\label{eq:elefin}\end{equation}

The $\beta$ derivatives are computed as follows

\begin{equation}
\begin{array}{c}
{\displaystyle \left\{ \frac{\partial^{\beta}F_{\beta k}\left(y,n,r\right)}{\partial y^{\beta}}\right\} _{y=0}=\sum_{L=0}^{\beta}\left(\begin{array}{c}
\beta\\
L\end{array}\right)\left\{ \frac{\partial^{L}}{\partial y^{L}}\left[\left[n-2k-r-1-y\right]^{2\beta+n-2}\right]\right\} _{y=0}}\\
{\displaystyle \times\left\{ \frac{\partial^{\beta-L}}{\partial y^{\beta-L}}\left[y+2r\right]^{-\beta-1}\right\} _{y=0},}\end{array}\label{eq:asd0}\end{equation}
to obtain

\begin{equation}
{\displaystyle \left\{ \frac{\partial^{\beta}F_{\beta k}\left(y,n,r\right)}{\partial y^{\beta}}\right\} _{y=0}=\sum_{L=0}^{\beta}\left(\begin{array}{c}
\beta\\
L\end{array}\right)\frac{\left(-\right)^{\beta}\left(2\beta-L\right)!\left(2\beta+n-2\right)!\left[n-2k-r-1\right]^{2\beta+n-2-L}}{\left(2\beta+n-2-L\right)!\left(\beta\right)!\left(2r\right)^{2\beta+1-L}}.}\label{eq:asd5}\end{equation}

Finally, we replace Eq. (\ref{eq:asd5}) into Eq. (\ref{eq:elefin})
and the resulting expression into Eq. (\ref{eq:imaginarypart}) to
obtain the final expression of $Im\left\{ _{3}F_{2}\right\} $ 

\begin{equation}
\begin{array}{c}
{\displaystyle Im\left\{ _{3}F_{2}\right\} =-\frac{\pi\Gamma\left(2\beta+n\right)}{\left(n-2k-1\right)^{2\beta+n-1}\left[\Gamma\left(\beta+1\right)\right]^{2}}\sum_{L=0}^{\beta}\left(\begin{array}{c}
\beta\\
L\end{array}\right)\frac{\left(2\beta-L\right)!}{\left(2\right)^{2\beta+1-L}}}\\
{\displaystyle \frac{r^{L}\left[n-2k-r-1\right]^{2\beta+n-2-L}}{\left(2\beta+n-2-L\right)!}\quad n\geq2}.\end{array}\label{eq:hyperim}\end{equation}

\clearpage\subsection*{List of Figures}

\begin{itemize}

\item[FIG. 1:] Normalized polymer propagator $4 \pi R^2 P \left( R, n , \alpha \right) / Z_n \left( \alpha \right) $ versus $R/n $ for $ n=5 $. Continuous line $\left( \alpha = 0.33 \right) $ , dotted line $\left( \alpha = 1.0 \right) $, dashed line $\left( \alpha = 3.0 \right) $, long dashed line $\left( \alpha = 5.0 \right) $ and dashed-dotted line $\left( \alpha = 10.0 \right) $.

\item[FIG. 2:] Normalized polymer propagator $4 \pi R^2 P \left( R, n , \alpha \right) / Z_n \left( \alpha \right) $ versus $R/n $ for $ n=10 $. Continuous line $\left( \alpha = 0.33 \right) $ , dotted line $\left( \alpha = 1.0 \right) $, dashed line $\left( \alpha = 3.0 \right) $, long dashed line $\left( \alpha = 5.0 \right) $ and dashed-dotted line $\left( \alpha = 10.0 \right) $.

\item[FIG. 3:] Normalized polymer propagator $4 \pi R^2 P \left( R, n , \alpha \right) / Z_n \left( \alpha \right) $ versus $R/n $ for $ n=30 $. Continuous line $\left( \alpha = 0.33 \right) $ , dotted line $\left( \alpha = 1.0 \right) $, dashed line $\left( \alpha = 3.0 \right) $, long dashed line $\left( \alpha = 5.0 \right) $ and dashed-dotted line $\left( \alpha = 10.0 \right) $.

\item[FIG. 4:] Normalized polymer propagator $4 \pi R^2 P \left( R, n , \alpha \right) / Z_n \left( \alpha \right) $ versus $R/n $ for $\alpha = 3.0 $. Continuous line $ \left( n =30 \right) $, dashed line $\left( n = 10 \right) $ and dashed-dotted line $ \left( n = 5 \right) $.

\item[FIG. 5:] Single chain structure factor $ S \left( k, n, \alpha \right) $ versus wave vector $ k $ for $n=5$. Continuous line $ \left( \alpha = 0.33 \right) $, dotted line $ \left( \alpha = 1.0 \right) $, dashed line $ \left( \alpha = 3.0 \right) $, dashed-dotted line $ \left( \alpha = 10.0 \right) $ and circles (best fits to the power law in the range $ k \in \left( 2 , 3 \right) $ for $ \alpha = 0.33 $  and $ \alpha = 10.0 $).

\item[FIG. 6:] Single chain structure factor $ S \left( k, n, \alpha \right) $ versus wave vector $ k $ for $n=10$. Continuous line $ \left( \alpha = 0.33 \right) $, dotted line $ \left( \alpha = 1.0 \right) $, dashed line $ \left( \alpha = 3.0 \right) $, dashed-dotted line $ \left( \alpha = 10.0 \right) $ and circles (best fits to the power law in the range $ k \in \left( 2 , 3 \right) $ for $ \alpha = 0.33 $  and $ \alpha = 10.0 $).

\item[FIG. 7:] Single chain structure factor $ S \left( k, n, \alpha \right) $ versus wave vector $ k $ for $n=30$. Continuous line $ \left( \alpha = 0.33 \right) $, dotted line $ \left( \alpha = 1.0 \right) $, dashed line $ \left( \alpha = 3.0 \right) $, dashed-dotted line $ \left( \alpha = 10.0 \right) $ and circles (best fits to the power law in the range $ k \in \left( 2 , 3 \right) $ for $ \alpha = 0.33 $  and $ \alpha = 10.0 $).
\item[FIG. 8:] Single chain structure factor $S\left(k,n,\alpha\right)$
versus wave vector $k$ for $n=30$. (Line) Kholodenko's model with $a=1$,
(points) this work with $\alpha = 2$.
\item[FIG. 9:] Single chain structure factor $S\left(k,n,\alpha\right)$
versus wave vector $k$ for $n=30$. (Line) Kholodenko's model with $a=50$,
(points) this work with $\alpha = 100$.
\item[FIG. 10:] Normalized polymer propagator $P\left(R,n,\alpha\right)$
versus end-to-end distance $R$ in units of Kuhn length for $n=5$. (Continuous lines) this work, (dashed lines) Wilhelm and Frey's results.

\end{itemize}\clearpage

Comment: Figure 1, First Author: Marcelo Marucho, Journal PRE

\begin{figure}[t]
\includegraphics[scale=0.5]{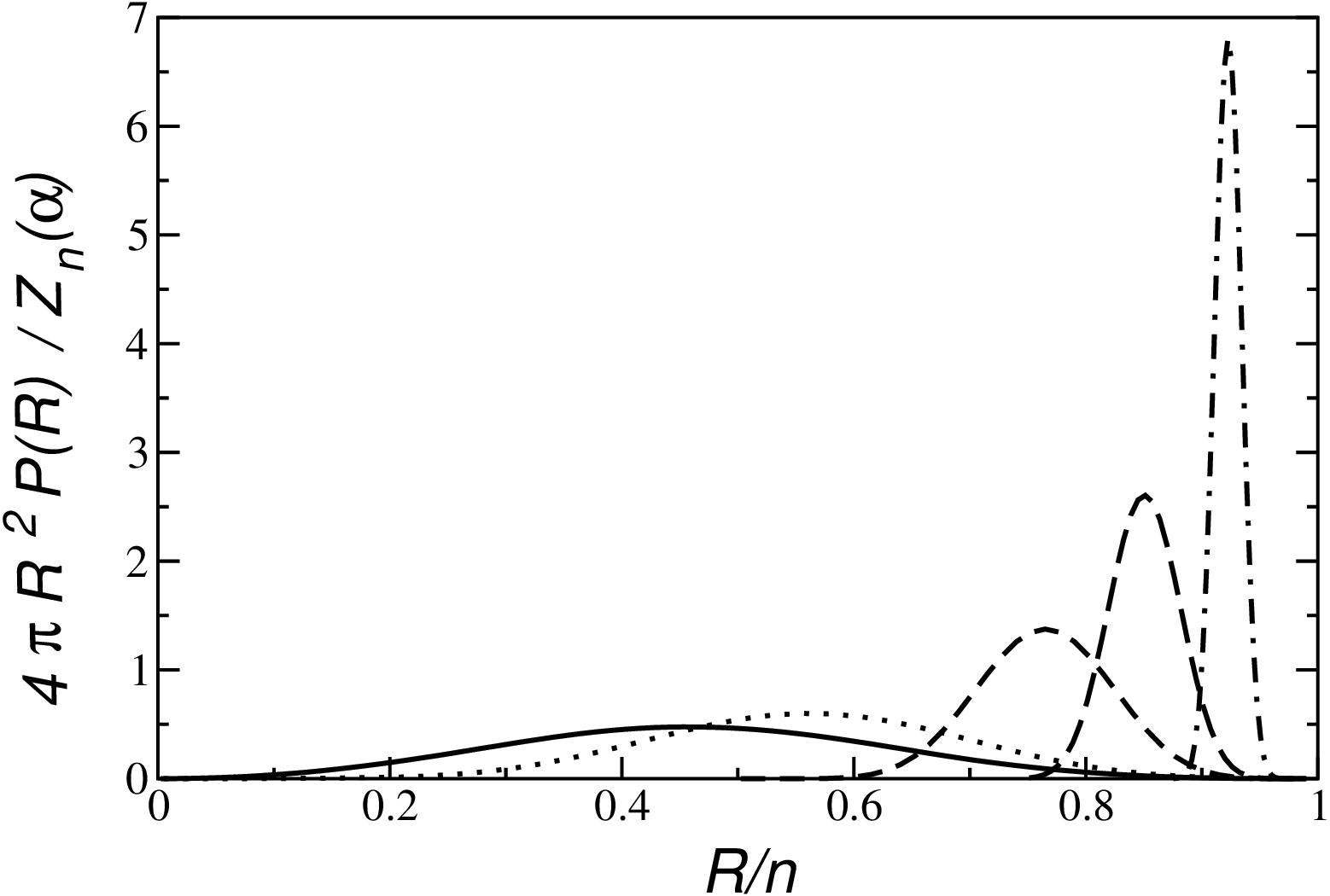}

FIG. 1. Normalized polymer propagator $4\pi R^{2}P\left(R,n,\alpha\right)/Z_{n}\left(\alpha\right)$
versus $R/n$ for $n=5$. Continuous line $\left(\alpha=0.33\right)$,
dotted line $\left(\alpha=1.0\right)$, dashed line $\left(\alpha=3.0\right)$,
long dashed line $\left(\alpha=5.0\right)$ and dashed-dotted line
$\left(\alpha=10.0\right)$.
\end{figure}

\clearpage

Comment: Figure 2, First Author: Marcelo Marucho, Journal PRE

\begin{figure}[t]
\includegraphics[scale=0.5]{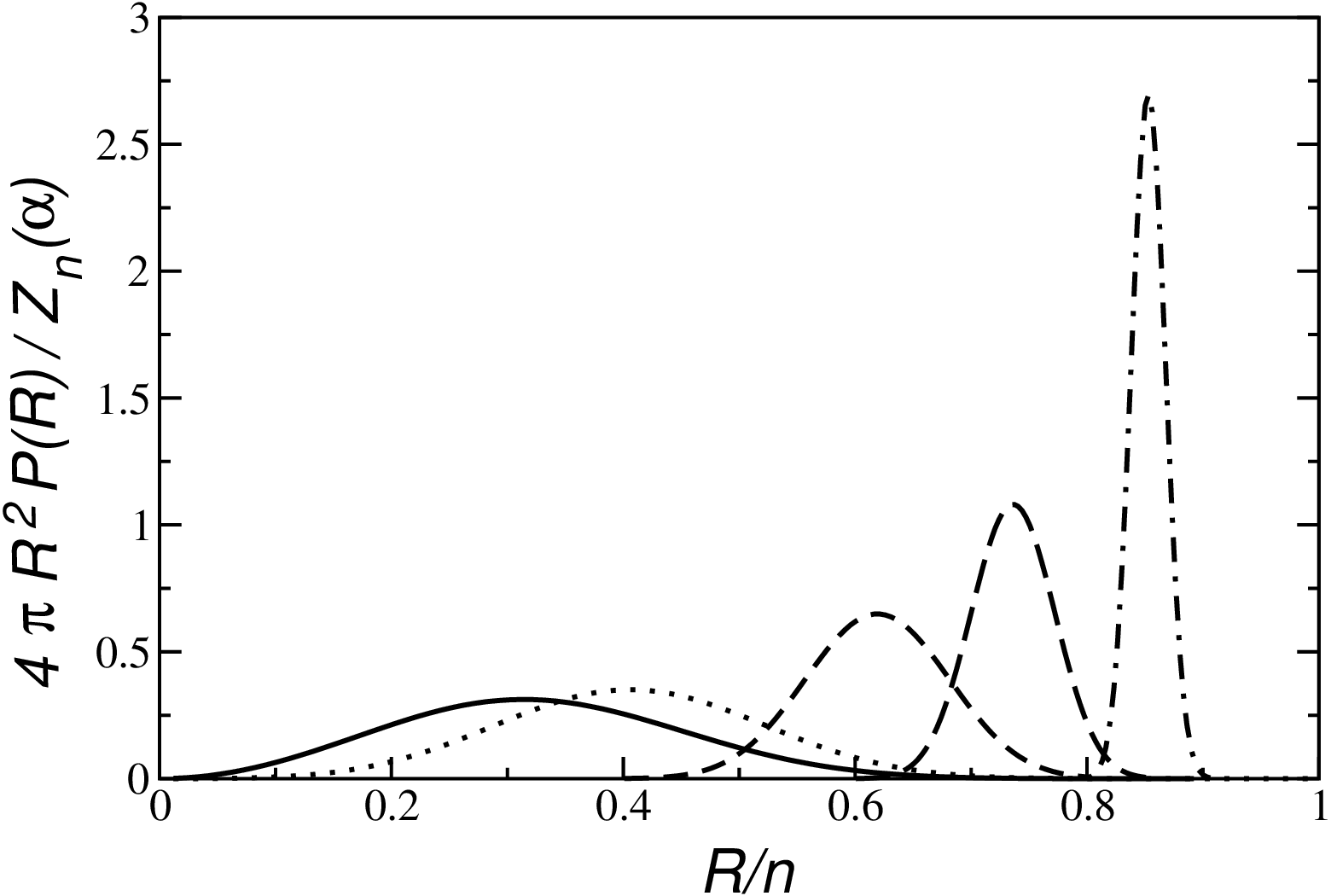}

FIG. 2. Normalized polymer propagator $4\pi R^{2}P\left(R,n,\alpha\right)/Z_{n}\left(\alpha\right)$
versus $R/n$ for $n=10$. Continuous line $\left(\alpha=0.33\right)$,
dotted line $\left(\alpha=1.0\right)$, dashed line $\left(\alpha=3.0\right)$,
long dashed line $\left(\alpha=5.0\right)$ and dashed-dotted line
$\left(\alpha=10.0\right)$.
\end{figure}

\clearpage

Comment: Figure 3, First Author: Marcelo Marucho, Journal PRE

\begin{figure}[t]
\includegraphics[scale=0.5]{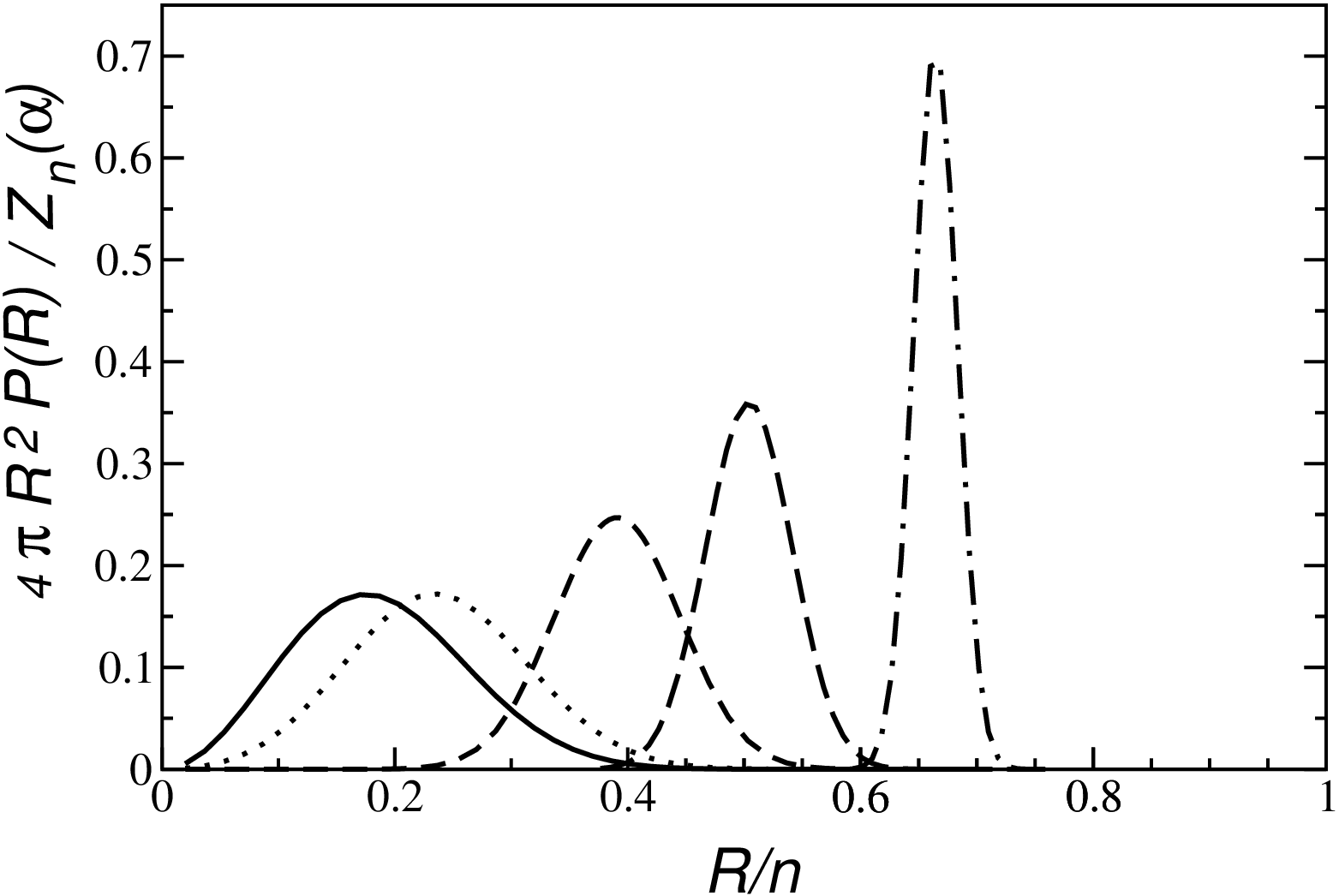}

FIG. 3. Normalized polymer propagator $4\pi R^{2}P\left(R,n,\alpha\right)/Z_{n}\left(\alpha\right)$
versus $R/n$ for $n=30$. Continuous line $\left(\alpha=0.33\right)$,
dotted line $\left(\alpha=1.0\right)$, dashed line $\left(\alpha=3.0\right)$,
long dashed line $\left(\alpha=5.0\right)$ and dashed-dotted line
$\left(\alpha=10.0\right)$.
\end{figure}

\clearpage

Comment: Figure 4, First Author: Marcelo Marucho, Journal PRE

\begin{figure}[t]
\includegraphics[scale=0.5]{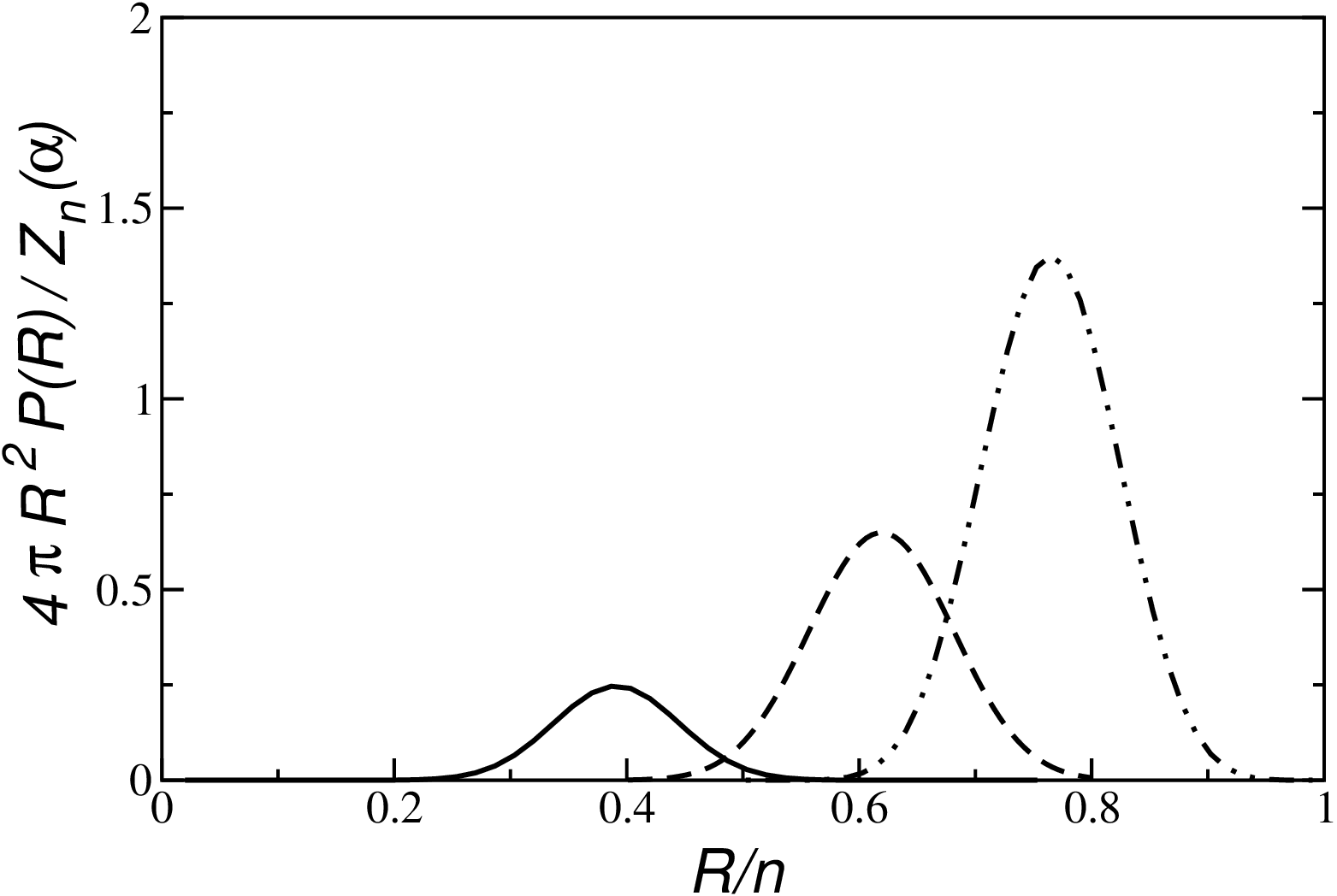}

FIG. 4. Normalized polymer propagator $4\pi R^{2}P\left(R,n,\alpha\right)/Z_{n}\left(\alpha\right)$
versus $R/n$ for $\alpha=3.0$. Continuous line $\left(n=30\right)$,
dashed line $\left(n=10\right)$ and dashed-dotted line $\left(n=5\right)$.
\end{figure}

\clearpage

Comment: Figure 5, First Author: Marcelo Marucho, Journal PRE

\begin{figure}[t]
\includegraphics[scale=0.5]{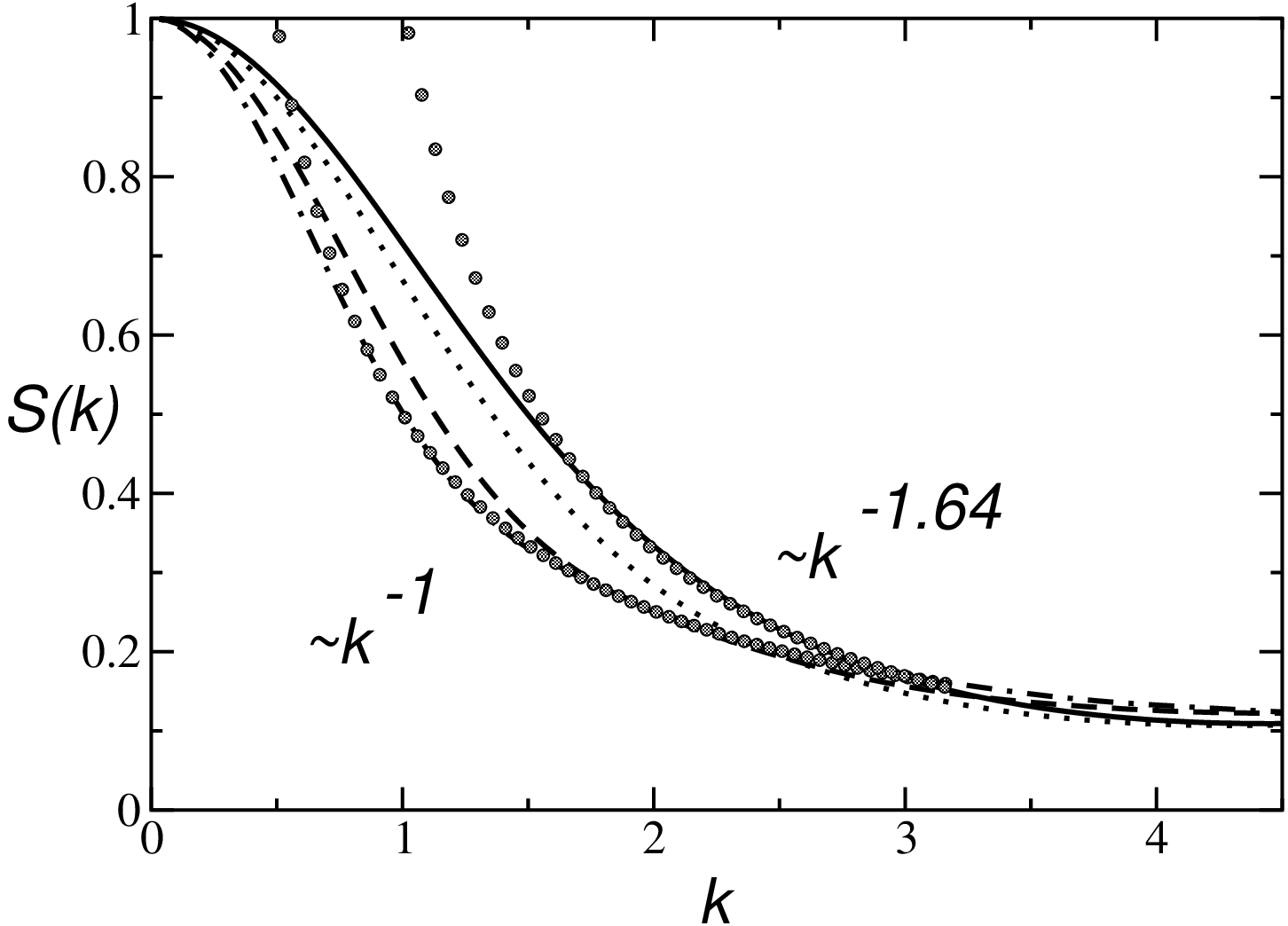}

FIG. 5. Single chain structure factor $S\left(k,n,\alpha\right)$
versus wave vector $k$ for $n=5$. Continuous line $\left(\alpha=0.33\right)$,
dotted line $\left(\alpha=1.0\right)$, dashed line $\left(\alpha=3.0\right)$,
dashed-dotted line $\left(\alpha=10.0\right)$ and circles (best fits
to the power law in the range $k\in\left(2,3\right)$ for $\alpha=0.33$
and $\alpha=10.0$).
\end{figure}

\clearpage

Comment: Figure 6, First Author: Marcelo Marucho, Journal PRE

\begin{figure}[t]
\includegraphics[scale=0.5]{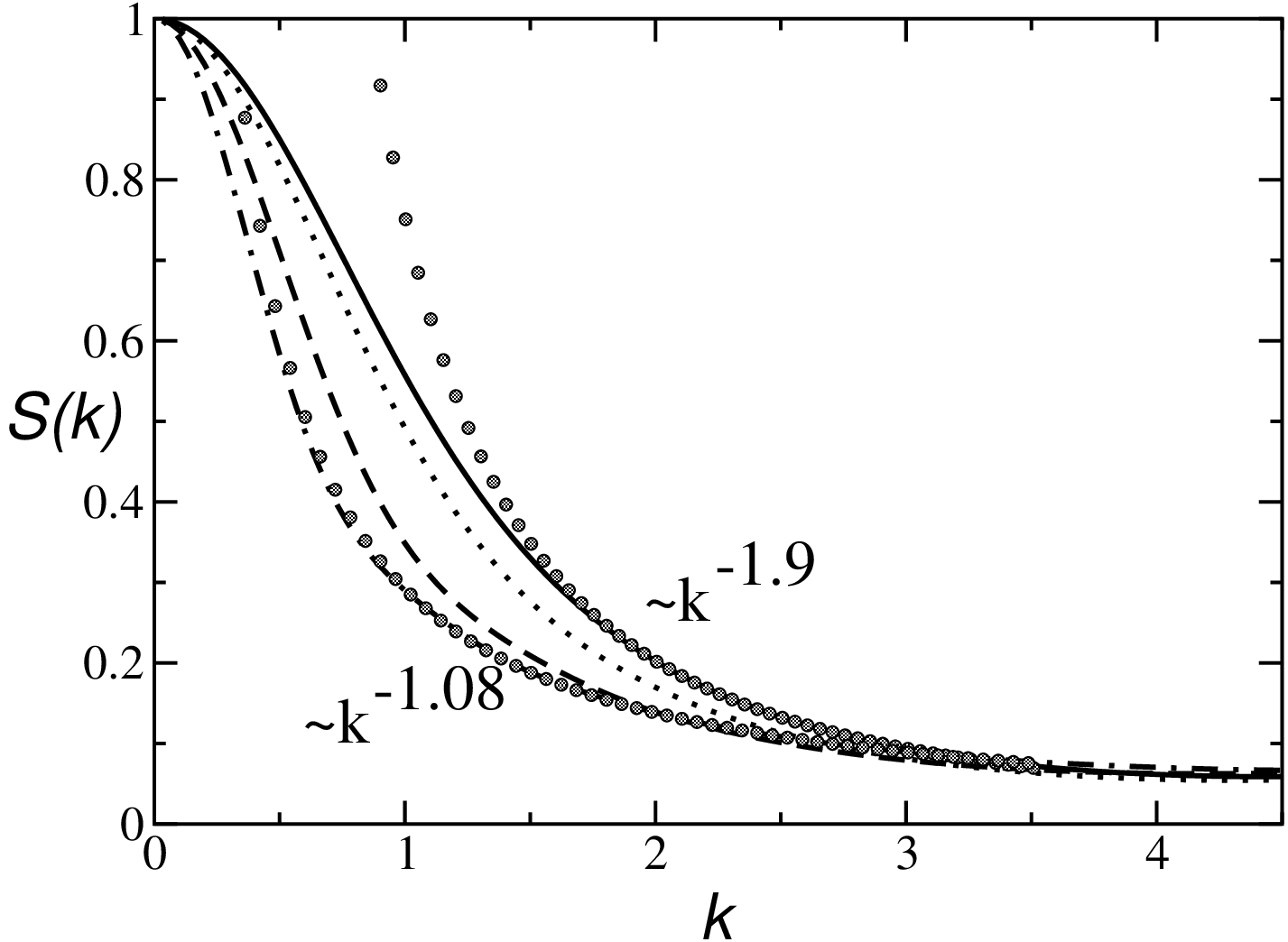}

FIG. 6. Single chain structure factor $S\left(k,n,\alpha\right)$
versus wave vector $k$ for $n=10$. Continuous line $\left(\alpha=0.33\right)$,
dotted line $\left(\alpha=1.0\right)$, dashed line $\left(\alpha=3.0\right)$,
dashed-dotted line $\left(\alpha=10.0\right)$ and circles (best fits
to the power law in the range $k\in\left(2,3\right)$ for $\alpha=0.33$
and $\alpha=10.0$).
\end{figure}

\clearpage

Comment: Figure 7, First Author: Marcelo Marucho, Journal PRE

\begin{figure}[t]
\includegraphics[scale=0.5]{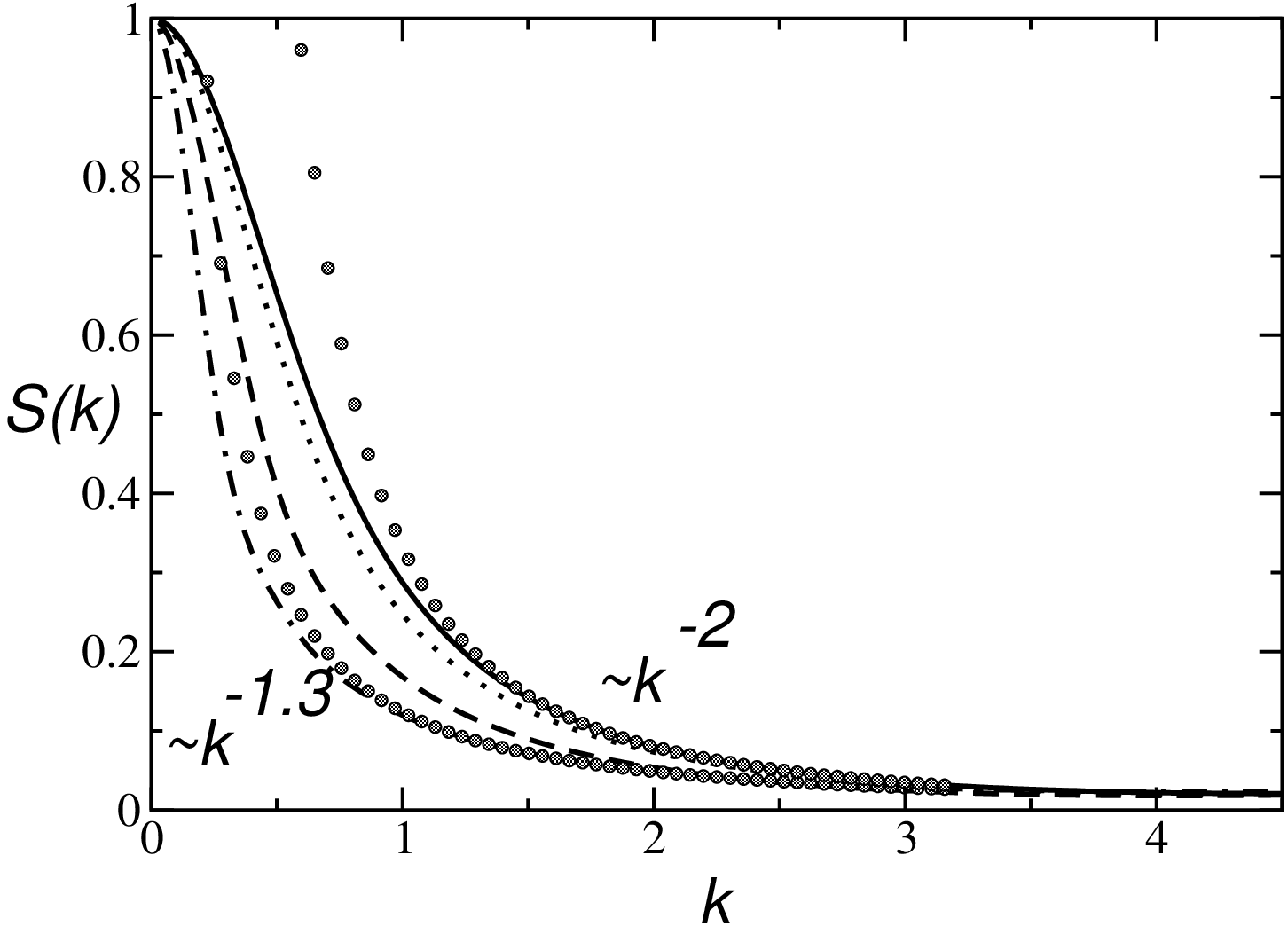}

FIG. 7. Single chain structure factor $S\left(k,n,\alpha\right)$
versus wave vector $k$ for $n=30$. Continuous line $\left(\alpha=0.33\right)$,
dotted line $\left(\alpha=1.0\right)$, dashed line $\left(\alpha=3.0\right)$,
dashed-dotted line $\left(\alpha=10.0\right)$ and circles (best fits
to the power law in the range $k\in\left(2,3\right)$ for $\alpha=0.33$
and $\alpha=10.0$).
\end{figure}

\clearpage

Comment: Figure 8, First Author: Marcelo Marucho, Journal PRE

\begin{figure}[t]
\includegraphics[scale=0.5]{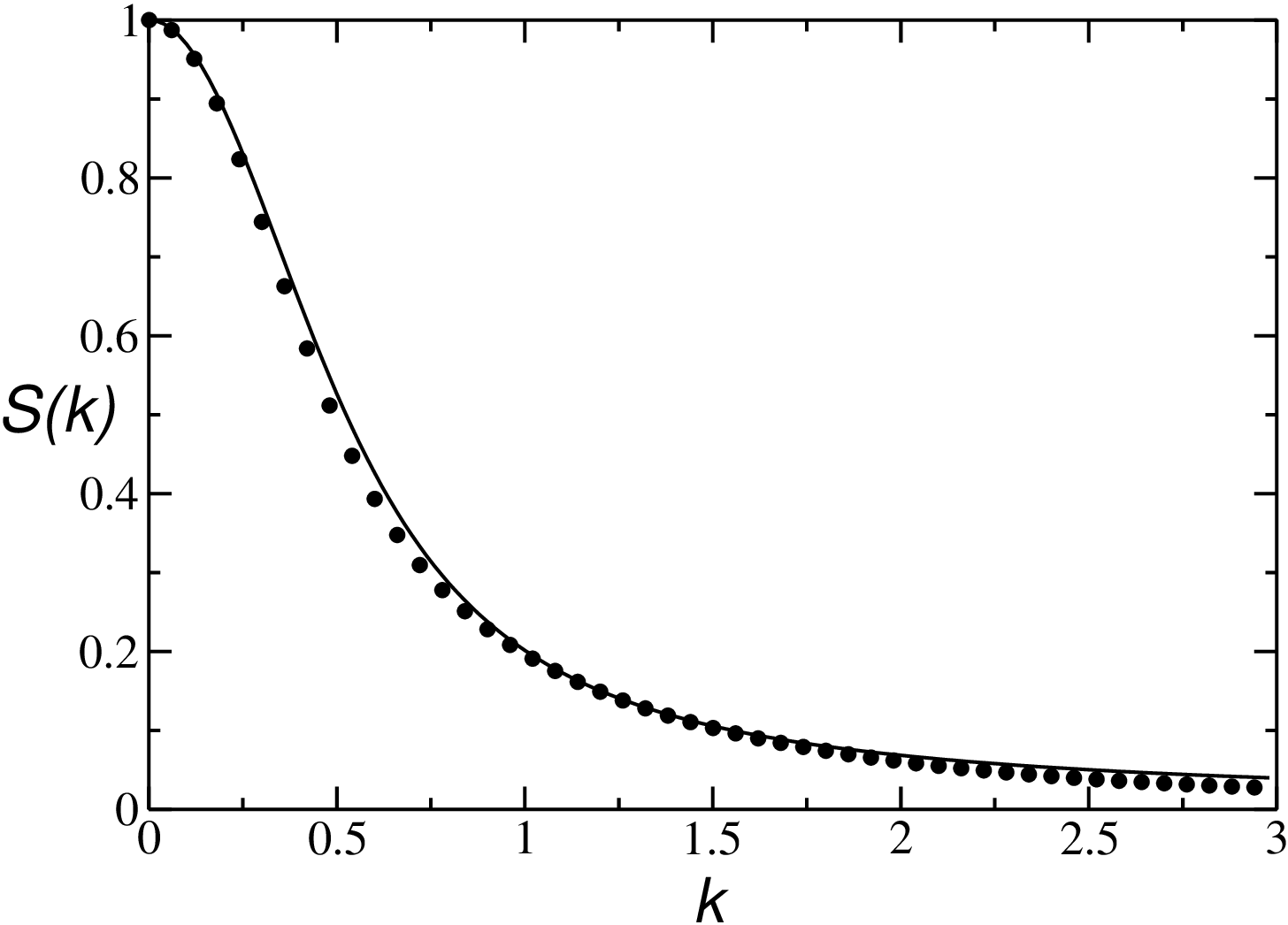}

FIG. 8. Single chain structure factor $S\left(k,n,\alpha\right)$
versus wave vector $k$ for $n=30$. (Line) Kholodenko's model with $a=1$,
(points) this work with $\alpha = 2$.
\end{figure}

\clearpage

Comment: Figure 9, First Author: Marcelo Marucho, Journal PRE

\begin{figure}[t]
\includegraphics[scale=0.5]{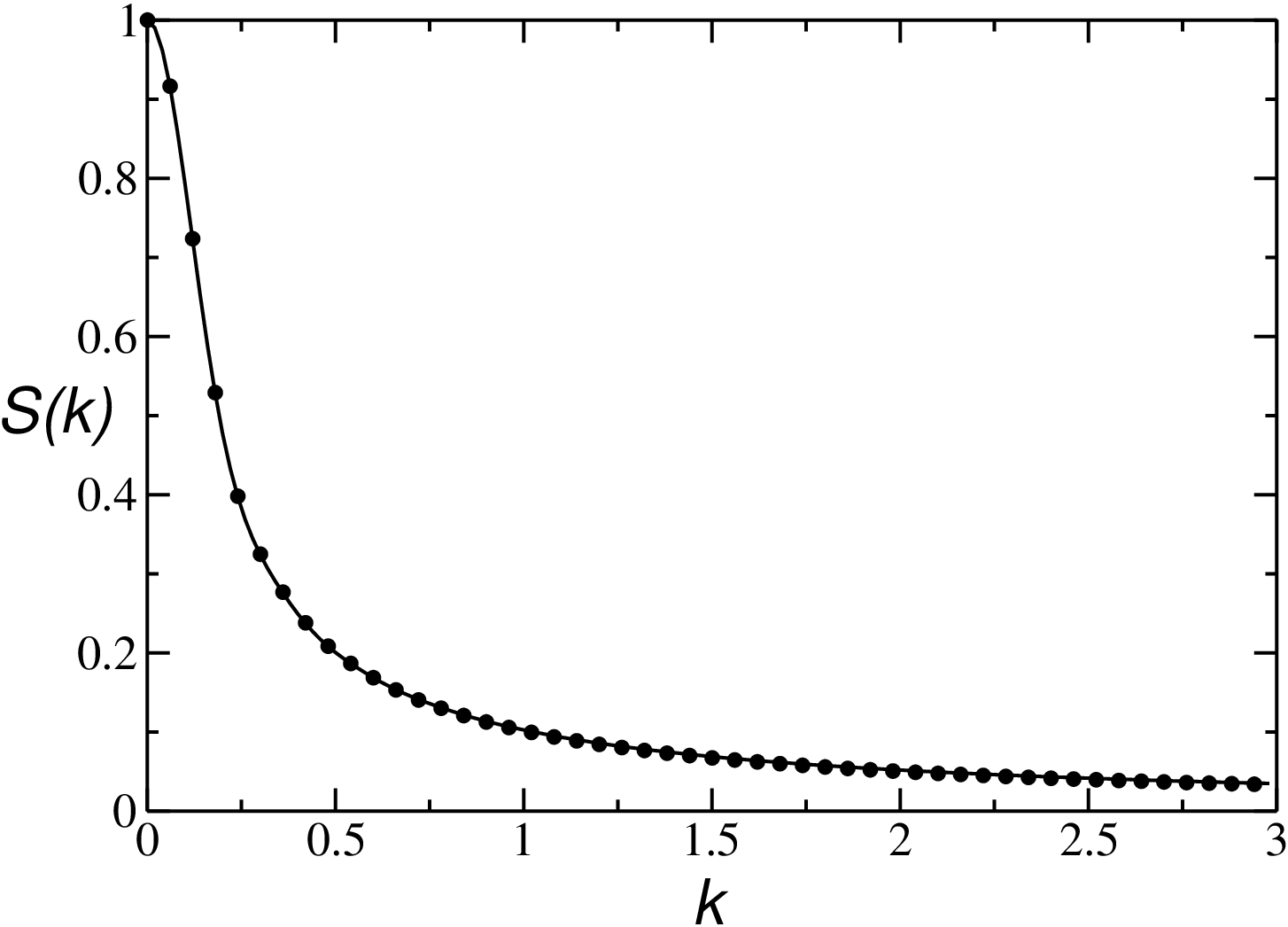}

FIG. 9. Single chain structure factor $S\left(k,n,\alpha\right)$
versus wave vector $k$ for $n=30$. (Line) Kholodenko's model with $a=50$,
(points) this work with $\alpha = 100$.
\end{figure}

\clearpage

Comment: Figure 10, First Author: Marcelo Marucho, Journal PRE

\begin{figure}[t]
\includegraphics[scale=0.5]{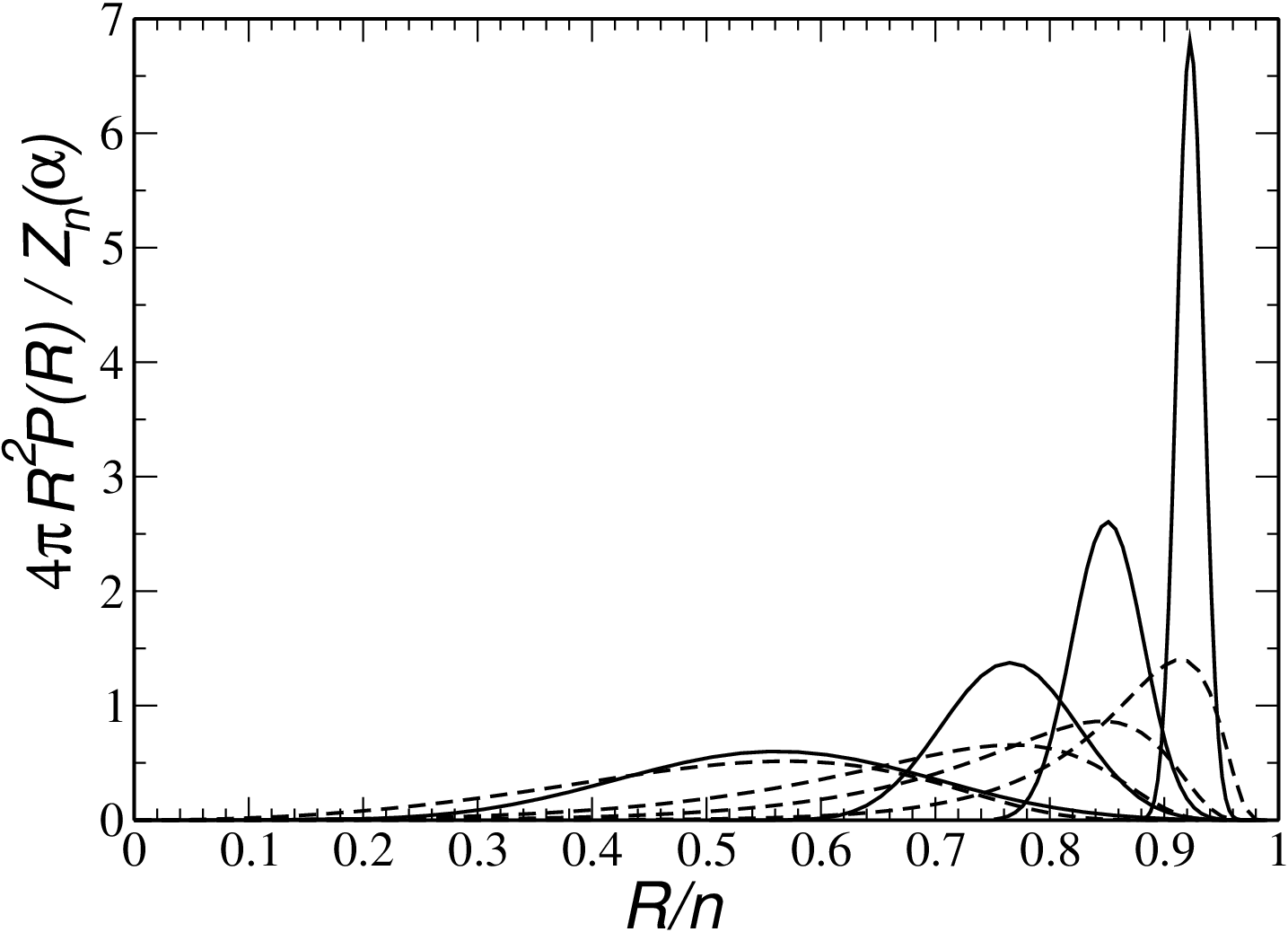}

FIG. 10. Normalized polymer propagator $P\left(R,n,\alpha\right)$
versus end-to-end distance $R$ in units of Kuhn length for $n=5$. (Continuous lines) this work, (dashed lines) Wilhelm and Frey's results.
\end{figure}

\end{document}